\documentstyle[12pt,indent,epsf,eqsection,subeqnarray]{article}
\begin{document}

\bibliographystyle{plain}

\date{March 31, 1999 \\[1mm] revised October 6, 1999}

\title{\vspace*{-1cm} Universal Amplitude Ratios in the \\
       Critical Two-Dimensional Ising Model \\
       on a Torus}

\author{
  {\small Jes\'us Salas}                                    \\[-0.2cm]
% {\small{\it Departamento de F\'{\i}sica de la Materia Condensada} and}
%                                                           \\[-0.2cm]
  {\small\it Departamento de F\'{\i}sica Te\'orica}         \\[-0.2cm]
  {\small\it Facultad de Ciencias, Universidad de Zaragoza} \\[-0.2cm]
  {\small\it Zaragoza 50009, SPAIN}                         \\[-0.2cm]
  {\small\tt JESUS@MELKWEG.UNIZAR.ES}                       \\[+2mm]
  {\small Alan D.~Sokal}                  \\[-0.2cm]
  {\small\it Department of Physics}       \\[-0.2cm]
  {\small\it New York University}         \\[-0.2cm]
  {\small\it 4 Washington Place}          \\[-0.2cm]
  {\small\it New York, NY 10003 USA}      \\[-0.2cm]
  {\small\tt SOKAL@NYU.EDU}               \\[-0.2cm]
  {\protect\makebox[5in]{\quad}}  % To force authors' names to be written
                                  %   vertically, one above another.
                                  % (\author seems to put them side-by-side
                                  %   if there is room.)
  \\
}
\vspace{0.5cm}

\maketitle
\thispagestyle{empty}   % Suppress page number on front page.

%\ltapprox and \gtapprox produce > and < signs with twiddle underneath
\def\spose#1{\hbox to 0pt{#1\hss}}
\def\ltapprox{\mathrel{\spose{\lower 3pt\hbox{$\mathchar"218$}}
 \raise 2.0pt\hbox{$\mathchar"13C$}}}
\def\gtapprox{\mathrel{\spose{\lower 3pt\hbox{$\mathchar"218$}}
 \raise 2.0pt\hbox{$\mathchar"13E$}}}
\def\inapprox{\mathrel{\spose{\lower 3pt\hbox{$\mathchar"218$}}
 \raise 2.0pt\hbox{$\mathchar"232$}}}

%%\doublespace

\begin{abstract}
Using results from conformal field theory,
we compute several universal amplitude ratios
for the two-dimensional Ising model at criticality on a symmetric torus.
These include the correlation-length ratio
$x^\star = \lim_{L\rightarrow\infty} \xi(L)/L$
and the first four magnetization moment ratios
$V_{2n} = \langle{\cal M}^{2n}\rangle/\langle{\cal M}^2\rangle^n$.
As a corollary we get the first four renormalized $2n$-point coupling 
constants for the massless theory on a symmetric torus, $G_{2n}^*$. 
We confirm these predictions by a high-precision Monte Carlo simulation.
\end{abstract}


%% \bigskip
%% \noindent
%% {\bf PACS Numbers:} 05.10.Cc, 05.50.+q, 05.70.Jk, 11.10.Kk, 11.25.Hf,
%%     64.60.Cn.

\clearpage

\newcommand{\be}{\begin{equation}}
\newcommand{\ee}{\end{equation}}
\newcommand{\<}{\langle}
\renewcommand{\>}{\rangle}
\newcommand{\para}{\|}
\renewcommand{\perp}{\bot}

\def\smfrac#1#2{{\textstyle\frac{#1}{#2}}}
\def\half{ {{1 \over 2 }}}
\def\smhalf{ {\smfrac{1}{2}} }
\def\scra{{\cal A}}
\def\scrc{{\cal C}}
\def\scrd{{\cal D}}
\def\scre{{\cal E}}
\def\scrf{{\cal F}}
\def\scrg{{\cal G}}
\def\scrh{{\cal H}}
\def\scrj{{\cal J}}
\def\scrk{{\cal K}}
\def\scrl{{\cal L}}
\def\scrm{{\cal M}}
\newcommand{\scrmvec}{\vec{\cal M}_V}
\def\scrmtens{{\stackrel{\leftrightarrow}{\cal M}_T}}
\def\scro{{\cal O}}
\def\scrp{{\cal P}}
\def\scrr{{\cal R}}
\def\scrs{{\cal S}}
\def\ttens{{\stackrel{\leftrightarrow}{T}}}
\def\scrv{{\cal V}}
\def\scrw{{\cal W}}
\def\scry{{\cal Y}}
\def\tauss{\tau_{int,\,\scrm^2}}
\def\taux{\tau_{int,\,{\cal M}^2}}
\newcommand{\taum}{\tau_{int,\,\vec{\cal M}}}
\def\taue{\tau_{int,\,{\cal E}}}
\newcommand{\imag}{\mathop{\rm Im}\nolimits}
\newcommand{\real}{\mathop{\rm Re}\nolimits}
\newcommand{\tr}{\mathop{\rm tr}\nolimits}
\newcommand{\sgn}{\mathop{\rm sgn}\nolimits}
\newcommand{\codim}{\mathop{\rm codim}\nolimits}
\newcommand{\rank}{\mathop{\rm rank}\nolimits}
\newcommand{\sech}{\mathop{\rm sech}\nolimits}
\def\textprime{{${}^\prime$}}
\newcommand{\longto}{\longrightarrow}
\def\var{ \hbox{var} }
\newcommand{\gtilde}{ {\widetilde{G}} }
\newcommand{\USp}{ \hbox{\it USp} }
\newcommand{\CP}{ \hbox{\it CP\/} }
\newcommand{\QP}{ \hbox{\it QP\/} }
\def\hboxscript#1{ {\hbox{\scriptsize\em #1}} }

\newcommand{\plotdot}{\makebox(0,0){$\bullet$}}
\newcommand{\plotsmalldot}{\makebox(0,0){{\footnotesize $\bullet$}}}

\def\bsigma{\mbox{\protect\boldmath $\sigma$}}
\def\bpi{\mbox{\protect\boldmath $\pi$}}
\def\btau{\mbox{\protect\boldmath $\tau$}}
  % \boldmath is fragile, and without the \protect we get screwed when
  % we try to use \bsigma in a \caption.
\def\bn{{\bf n}}
\def\br{{\bf r}}
\def\bz{{\bf z}}
\def\bh{\mbox{\protect\boldmath $h$}}

\def\betatilde{ {\widetilde{\beta}} }
\def\hatp{\hat p}
\def\hatl{\hat l}

\def\msbar{ {\overline{\hbox{\scriptsize MS}}} }
\def\normalmsbar{ {\overline{\hbox{\normalsize MS}}} }

\def\eff{ {\hbox{\scriptsize\em eff}} }

\newcommand{\reff}[1]{(\ref{#1})}

%\font\specialroman=msym10 scaled\magstep1 % 12-point Special Roman (caps only)
%\font\sevenspecialroman=msym7             % 7-point Special Roman (caps only)
\def\Z{\hbox{\specialroman Z}}
\def\zed{\hbox{\specialroman Z}}
\def\szed{\hbox{\sevenspecialroman Z}}
\def\R{\hbox{{\bf R}}}
\def\sR{\hbox{\sevenspecialroman R}}
\def\N{\hbox{\specialroman N}}
\def\C{\hbox{\specialroman C}}
\def\Q{\hbox{\specialroman Q}}
\renewcommand{\emptyset}{\hbox{\specialroman ?}}
%\newcommand{\Z}{{\bf Z}}
%\newcommand{\zed}{{\bf \Z}}
%\newcommand{\R}{\hbox{{\rm I}\kern-.2em\hbox{\rm R}}}
%\font\srm=cmr7                 % to get seven roman
%\def\szed{\hbox{\srm Z\kern-.45em\hbox{\srm Z}}}
%\def\sR{\hbox{{\srm I}\kern-.2em\hbox{\srm R}}}
%\def\C{{\bf C}}

% \font\german=eufm10 scaled\magstep1   % 12-point Euler Fraktur (German)
% \def\germang{\hbox{\german g}}
% \def\germansu{\hbox{\german su}}

% \font\amssymbol=msxm10 scaled \magstep1  % Another AMS symbol font
% \def\transversal{\hbox{\amssymbol t}}  % THERE MAY BE A BETTER SYMBOL.

\newtheorem{theorem}{Theorem}[section]
\newtheorem{corollary}[theorem]{Corollary}
\newtheorem{lemma}[theorem]{Lemma}
\newtheorem{conjecture}[theorem]{Conjecture}
\newtheorem{definition}[theorem]{Definition}
\def\proof{\bigskip\par\noindent{\sc Proof.\ }}
\def\qed{\hbox{\hskip 6pt\vrule width6pt height7pt depth1pt \hskip1pt}\bigskip}

%
% Array for subscripts
%
\newenvironment{sarray}{
          \textfont0=\scriptfont0
          \scriptfont0=\scriptscriptfont0
          \textfont1=\scriptfont1
          \scriptfont1=\scriptscriptfont1
          \textfont2=\scriptfont2
          \scriptfont2=\scriptscriptfont2
          \textfont3=\scriptfont3
          \scriptfont3=\scriptscriptfont3
        \renewcommand{\arraystretch}{0.7}
        \begin{array}{l}}{\end{array}}

\newenvironment{scarray}{
          \textfont0=\scriptfont0
          \scriptfont0=\scriptscriptfont0
          \textfont1=\scriptfont1
          \scriptfont1=\scriptscriptfont1
          \textfont2=\scriptfont2
          \scriptfont2=\scriptscriptfont2
          \textfont3=\scriptfont3
          \scriptfont3=\scriptscriptfont3
        \renewcommand{\arraystretch}{0.7}
        \begin{array}{c}}{\end{array}}

%%%%%%%%%%%%% BEGINNING OF THE TEXT %%%%%%%%%%%%%%%%%%%%%%%%%%%%%%%%%%%%

\section{Introduction}   \label{sec_intro}

A central concept in the theory of critical phenomena
is the idea of {\em universality}\/,
which states that phase-transition systems can be divided
into a relatively small number of ``universality classes''
(determined primarily by the system's spatial dimensionality
and the symmetries of its order parameter)
within which certain features of critical behavior are universal.
In the 1950s and 1960s it came to be understood that
critical exponents are universal in this sense \cite{Domb_96}.
Later, in the 1970s,
it was learned that
certain dimensionless ratios of critical amplitudes are also universal
\cite{Privman_91}.

The past quarter-century has seen enormous progress in the determination
of critical exponents for a wide variety of universality classes,
including exact analytical results for two-dimensional (2D) models
\cite{Nienhuis_87,Cardy_87,Itzykson_88,DiFrancesco_97}
and increasingly precise numerical determinations for three-dimensional models
by a variety of techniques
(field-theoretic renormalization group \cite{Zinn_Justin_book,Guida_98},
series extrapolation \cite{Bhanot_92,Janke_93,Guttmann_93,Guttmann_94,%
Arisue_95,Butera_97a,Butera_97b,Butera_99},
Monte Carlo \cite{Holm_93,Gottlob_93,Li_95,Bloete_95,Nightingale_96,Bloete_96,%
Caracciolo_98,Ballesteros_98,Ballesteros_99,Tsypin_99}).
As a result, attention has turned quite naturally to
universal amplitude ratios:
these include amplitude ratios in infinite volume
and those in finite-size scaling (FSS).
Though much numerical work has been done,
few exact results are known.\footnote{
   Among the models studied are
   the 2D Ising model
   \cite{Zinn_96b,Pelissetto_98,Pelissetto_98a,Sokolov_98,Kim_99},
   2D nonlinear $\sigma$-models
   \cite{CEPS_o3_prl,CEMPS_LAT95_4loop,mgsu3,Pelissetto_98,Pelissetto_98a,%
Balog_97,Balog_99},
   2D Potts models \cite{Caselle_99,Delfino_99},
   the Baxter 8-vertex model \cite{Naon_89},
   2D and 3D self-avoiding walks \cite{Li_95},
   the 3D Ising model
   \cite{Liu_Fisher_89,Kim-Patrascioiu_93,Bloete_95,Gutsfeld_96,%
Zinn_Fisher_96,Zinn_96b,Guida_Zinn-Justin_97,Caselle_Hasenbusch_97,%
Pelissetto_98,Pelissetto_98a,Pelissetto_98b,Ballesteros_98,Ballesteros_99,%
Kim_99,Tsypin_99},
   3D $O(N)$ spin models
   \cite{Gottlob_93,Janke_98,Butera_98,Pelissetto_98,Pelissetto_98a,Butera_99,%
Sokolov_99,Petkou_99},
   3D site percolation \cite{Ballesteros_99},
   and the 5D Ising model \cite{Luitjen_99}.
   This list of references is far from exhaustive.
}

The critical behavior of many 2D models
can be studied analytically using conformal field theory (CFT)
\cite{Cardy_87,Itzykson_88,DiFrancesco_97}.
Many critical exponents have been determined exactly,
along with a few universal amplitude ratios
\cite{Wu_76,Di_Francesco_87,Di_Francesco_88,Cardy_88a,Cardy_88b,%
Cardy_Saleur_89,Caracciolo_90,Cardy_Mussardo_93,Cardy_Guttmann_93,Smilga_97,%
Delfino_98,Delfino_Cardy_98,Delfino_99}.
The main goal of the present paper is to compute, using CFT,
a few more universal amplitude ratios for the 2D Ising model
and to test these predictions by a high-precision Monte Carlo study.
The amplitude ratios considered here arise in finite-size scaling;
they can be computed starting from the correlation functions
of the critical 2D Ising model on a torus.

The first class of quantities we study concern
the shape of the magnetization distribution $\rho({\cal M})$
at criticality on a symmetric torus ($L_x = L_y$).\footnote{
   One of the insights of conformal field theory is that
   universal finite-size-scaling properties
   (such as the universal amplitude ratios considered here)
   ought to be studied as analytic functions of the
   modular parameter $\tau$ of the torus.
   Nevertheless, we think that the case of a symmetric torus ($\tau = i$)
   is of sufficient practical importance in Monte Carlo simulations
   to warrant special attention.
}
We study the rescaled shape of this distribution
(i.e.\ normalizing by its width $\< {\cal M}^2 \>^{1/2}$)
as well as the dimensionless ratios of its moments,
\be
V_{2n}  \;=\; { \<{\cal M}^{2n}\> \over \<{\cal M}^2\>^n }   \;.
\label{def_ratios}
\ee
We can also define the dimensionless cumulants
\be
U_{2n} \equiv { \<{\cal M}^{2n}\>_{\rm conn} \over \<{\cal M}^2\>^n }   \;.
\label{def_ratios_bis}
\ee
For any symmetric distribution $\rho({\cal M}) = \rho(-{\cal M})$
these satisfy\footnote{
   These relations can be computed from the generating functions
   $$
   \sum\limits_{n=1}^\infty {U_{2n} \over (2n)!} z^{2n}   \;=\;
   \log\!\left( \sum\limits_{n=0}^\infty {V_{2n} \over (2n)!} z^{2n} \right)
   $$
   with $V_0 = V_2 = 1$ and $V_{2n+1} = 0$.
}
\begin{subeqnarray}
\slabel{def_U4}
U_4  &=& V_4 - 3 \\
U_6  &=& V_6 - 15 V_4 + 30 \\
U_8  &=& V_8 - 28 V_6 - 35 V_4^2 + 420 V_4  - 630  \\
U_{10} &=& V_{10} - 45 V_8 - 210 V_4 V_6 + 1260 V_6 + 3150 V_4^2
            - 18900 V_4 + 22680  \\
      & \vdots &   \nonumber
\end{subeqnarray}
Note that $V_4$ and $U_4$ are closely related to the
so-called Binder cumulant \cite{Binder_81}
\be
  U_{\rm 4,Binder} \equiv 1 - {\<{\cal M}^{4}\>\over 3 \<{\cal M}^2\>^2} =
                    1 - {V_4 \over 3} = - {U_4 \over 3}
  \;.
\label{def_Binder_cumulant}
\ee
For $\beta<\beta_c$ the ratios $V_{2n}$ tend in the infinite-volume limit
to those characteristic of a Gaussian distribution,
\begin{subeqnarray}
V_{2n}({\rm Gaussian})  & = & (2n-1)!!   \slabel{V2n_Gaussian} \\
U_{2n}({\rm Gaussian})  & = & 0
\label{ratios_Gaussian}
\end{subeqnarray}
while for $\beta>\beta_c$ they tend to those characteristic of
a sum of two delta functions,
\begin{subeqnarray}
V_{2n}(\mbox{\rm two deltas}) & = & 1   \slabel{V2n_2deltas} \\
U_{2n}(\mbox{\rm two deltas}) & = & {2^{2n-1} (2^{2n} - 1) \over n} \, B_{2n}
\label{ratios_2deltas}
\end{subeqnarray}
where $B_{2n} = (-1)^{n-1} (2n)! \, \zeta(2n) / 2^{2n-1} \pi^{2n}$
is a Bernoulli number.
At $\beta=\beta_c$, however, these ratios acquire non-trivial values
in-between \reff{V2n_Gaussian} and \reff{V2n_2deltas}.\footnote{
   The Schwarz inequality implies that $V_{2n} \ge 1$ for any model,
   and the Gaussian inequality \cite{Newman_75a,Bricmont_77}
   implies that $V_{2n} \le (2n-1)!!$ for ferromagnetic Ising models.
   In particular, we have $-2 \le U_4 \le 0$.
   Moreover, Newman \cite{Newman_75b} and Shlosman \cite{Shlosman_86}
   have proven, for ferromagnetic Ising models, that
   $(-1)^{n-1} U_{2n} \ge 0$ for all $n$;
   and Newman \cite{Newman_75b} has proven some additional inequalities
   on the $U_{2n}$.
}
These values, which are universal, can in principle be computed
by integrating the spin correlators
for the critical 2D Ising model on a torus,
which were determined by Di Francesco {\em et al.}\/
\cite{Di_Francesco_87,Di_Francesco_88}
using CFT.
In practice, however, the formula for $V_{2n}$
rapidly gets more complicated as $n$ grows.
Di Francesco {\em et al.}\/ \cite{Di_Francesco_87,Di_Francesco_88}
computed $V_4$ to roughly three decimal places by Monte Carlo integration.
Here we shall improve this result by three orders of magnitude,
and shall also compute $V_6$ to five decimal places,
$V_8$ to almost four decimal places, and $V_{10}$ to three decimal places:
\begin{eqnarray}
V_4     & = & 1.1679229 \pm 0.0000047    \\
V_6     & = & 1.4556491 \pm 0.0000072    \\
V_8     & = & 1.89252   \pm 0.00018      \\
V_{10}  & = & 2.53956   \pm 0.00034
\end{eqnarray}
Finally, we shall measure the ratios $V_{2n}$ for $2 \le n \le 10$
by Monte Carlo simulation, with an accuracy that gradually
deteriorates as $n$ grows.
For $n=2,3,4,5$ our Monte Carlo estimates agree well with
the theoretical predictions (but are of course less precise).

Another interesting quantity is the second-moment correlation
length $\xi$.
It has a sensible definition in finite volume (see Section \ref{sec_simul}),
and its expected FSS behavior is
\be
   \xi \;\sim\; L \left[ x^\star  + A L^{-\Delta} + \cdots \right]   \;,
 \label{def_FSS_xi}
\ee
where the leading coefficient
\be
x^\star  \;=\; \lim_{L\rightarrow\infty} \xi/L
\ee
is universal.\footnote{
   The quantity $x^\star$ also plays an important role
   in a recently developed method for extrapolating finite-volume
   Monte Carlo data to infinite volume
   \cite{Luscher_91,Kim_93,fss_greedy,mgsu3,fss_greedy_fullpaper}.
}${}^,$\footnote{
  An analogous situation holds in a cylindrical ($L \times \infty$) geometry
  for the exponential correlation length in the longitudinal direction,
  $\xi_{exp}(L)$
  [which can be defined in terms of the logarithm of
  the ratio of the two largest eigenvalues of the transfer matrix].
  Privman and Fisher \protect\cite{Privman_Fisher_84} showed that
  $\lim\limits_{L \to\infty} \xi_{exp}(L)/L$ at criticality
  is universal, and Cardy \protect\cite{Cardy_84} showed that
  for 2D conformal-invariant systems
  it is equal to $1/(\pi \eta)$.
}
Here we shall compute $x^\star$ for the 2D Ising model at criticality
by numerically integrating the known spin correlators
\cite{Di_Francesco_87,Di_Francesco_88};
we find
\be
x^\star \;=\; 0.90504 88292 \pm 0.00000 00004   \;.
\ee
Our Monte Carlo data confirm this prediction.
To our knowledge, this is the first exact determination of $x^\star$
for any universality class.
Monte Carlo estimates of $x^\star$
are available for many other 2D models --- including
the 3-state Potts model \cite{Salas_Sokal_Potts3},
the 4-state Potts model \cite{Salas_Sokal_AT,Salas_Sokal_FSS},
the 3-state square-lattice Potts antiferromagnet
\cite{Salas_Sokal_swaf3,Ferreira_Sokal_prep},
the XY model \cite{Kim_private},
and several points on the self-dual curve of the symmetric Ashkin--Teller model
  \cite{Salas_Sokal_AT}
--- and it would be very interesting to compute $x^\star$ analytically
for some of these models.
Numerical estimates of $x^\star$ are also available
for some three-dimensional spin models
\cite{Gottlob_93,Kim-Patrascioiu_93,Janke_98,Ballesteros_98}.

Consider, finally, the dimensionless renormalized four-point coupling constant
\be
   g_4 \;=\; - \, {\bar{u}_4 \over \chi^2 \xi^d}
       \;=\; - \, {U_4 \over (\xi/L)^d}
   \;,
\ee
where $\bar{u}_4$ is the connected four-point function at zero momentum.
In the FSS limit
$L \to \infty$, $\beta \to \beta_c$ with $\xi/L$ fixed,
$g_4$ is a nontrivial function of the FSS variable $\xi/L$:
\be
   g_4 \;=\; F_{g_4}(\xi/L)   \;.
\ee
Therefore, the function $g_4(\beta,L)$ fails to be
jointly continuous at $(\beta,L) = (\beta_c,\infty)$;
many limiting values are possible depending on the mode of approach,
and the massive and massless scaling limits
\begin{eqnarray}
   g_4^*  & = & \lim\limits_{\beta \uparrow \beta_c}
                \lim\limits_{L \to \infty}
                g_4(\beta,L)   \\[1mm]
   G_4^*  & = & \lim\limits_{L \to \infty}
                \lim\limits_{\beta \uparrow \beta_c}
                g_4(\beta,L)
        \;=\; \lim\limits_{L \to \infty}
                g_4(\beta_c,L)
\end{eqnarray}
correspond to the two extreme cases
$g_4^* = F_{g_4}(0)$, $G_4^* = F_{g_4}(x^\star)$.
As a corollary of our computation of $V_4$ and $x^\star$,
we obtain the value of $g_4$ at criticality on a symmetric torus:
\be
   G_4^*  \;=\;  - \, {U_4  \over  x^{\star 2}}
          \;=\;  2.2366587  \pm 0.0000057   \;.
\ee

More generally, consider the dimensionless renormalized $2n$-point
coupling constant
\be
   g_{2n} \;=\;  {\chi^n \, \bar{\Gamma}_{2n} \over \xi^{(n-1)d}}
   \;,
\ee
where $\bar{\Gamma}_{2n}$ is the amputated one-particle-irreducible
$2n$-point function at zero mo\-men\-tum.\footnote{
   The $\bar{\Gamma}_{2n}$ are defined by the generating-function relation
   $$
   \sum\limits_{n=1}^\infty {\bar{\Gamma}_{2n} \over (2n)!} \phi^{2n}   \;=\;
   \sup_J \left[ J\phi \,-\,
                 \sum\limits_{n=1}^\infty {\bar{u}_{2n} \over (2n)!} J^{2n}
          \right]
          \;.
   $$
   Recall that $\bar{u}_2 = \chi$ and that
   $\bar{u}_{2n} = L^{-d} \<{\cal M}^{2n}\>_{\rm conn}$.
   %
   % (* A Mathematica program to compute this *)
   %
   %  WW = Sum[u[2n] J^(2n) / (2n)!, {n,1,10}] + O[J]^22
   %  
   %  myJ = InverseSeries[D[WW,J], phi]
   %  
   %  myGamma = myJ*phi - Sum[u[2n] myJ^(2n) / (2n)!, {n,1,10}]
   %  
   %  Do[mygamma[2n] = Together[Coefficient[(2n)! * myGamma, phi^(2n)]],
   %     {n,1,10}]
   %
}
We can predict the next three renormalized coupling constants
at criticality on a symmetric torus:
\begin{subeqnarray}
G_6^* &=& - \, {U_6 - 10 U_4^2 \over x^{\star 4}}
                                          \;=\;  29.25457 \pm 0.00015 \\[1mm]
G_8^* &=& - \, {U_8 - 56 U_4 U_6 + 280 U_4^3  \over x^{\star 6}}
                                             \;=\;  942.6095  \pm 0.0072\\[1mm]
G_{10}^* &=& - \, {U_{10} - 120 U_4 U_8 - 126 U_6^2 + 4620 U_4^2 U_6
                     - 15400 U_4^4   \over x^{\star 8}} \nonumber \\ 
         & &  \hspace{4.5cm} =\; 56110.24  \pm 0.56 
\end{subeqnarray} 
In addition, we shall provide Monte Carlo estimates of
$G_{12}^*$ through $G_{20}^*$ [cf.\ \reff{G_2n_MC_results} below].

This paper is organized as follows:
In Section~\ref{sec_theor} we review the relevant exact results
available for the 2D Ising model at criticality on a torus,
and we compute (by numerical integration)
the CFT prediction for the quantities
$x^\star$, $V_4$, $V_6$, $V_8$ and $V_{10}$.
In Section~\ref{sec_simul} we explain the Monte Carlo algorithm we
have used to simulate this model.
In Section~\ref{sec_res_static} we analyze our numerical results
for the static observables and compare them against
the available exact results.
Finally, in Section~\ref{sec_conclusions} we present our final conclusions
and discuss prospects for future work.
In Appendix~\ref{app_integrals} we explain how we carried out
the numerical integrations involved in computing $x^\star$,
and in Appendix~\ref{app_theta} we summarize the definitions and
principal properties of the Jacobi theta functions.

\section{Theoretical Results}   \label{sec_theor}

The universal amplitudes we consider in this paper ($x^\star$ and $V_{2n}$) 
can be written in terms of integrals of the
$2n$-point spin correlation functions of the
the critical 2D Ising continuum field theory on a torus.
These correlators were obtained by Di Francesco {\em et al}.\
\cite{Di_Francesco_87,Di_Francesco_88} using
an approach based on conformal field theory.
The result is\footnote{
   There is a misprint in the normalization of the 4-spin correlator
   in equation (9) of \protect\cite{Di_Francesco_88}, and in
   the normalization of the $2n$-spin correlator in equation (6.6) of
   \protect\cite{Di_Francesco_87}.
   We have rederived both correlators
   using the chiral bosonization prescription presented in
   \protect\cite{Di_Francesco_87}. With the correct normalization,
   shown in
   (\protect\ref{def_ising_spin_correlators})--%
(\protect\ref{def_ising_spin_correlators_full})
    below,
   we are able to reproduce the numerical value of $V_4$
   reported in \protect\cite{Di_Francesco_88},
   as well as the numerical estimates
   of $V_4$, $V_6$, $V_8$ and $V_{10}$ obtained in our simulation.
}
\be
   \<\sigma_{z_1}\cdots\sigma_{z_{2n}}\>
   \;=\;
   {\sum_{\nu=1}^4 Z_\nu \<\sigma_{z_1}\cdots\sigma_{z_{2n}}\>_\nu
    \over
    \sum_{\nu=1}^4 Z_\nu
   }
 \label{def_ising_spin_correlators}
\ee
where
\be
   Z_\nu   \;=\;   {|\theta_\nu(0)|   \over   2 |\eta|}
 \label{def_ising_partition}
\ee
and
\begin{subeqnarray}
 Z^2_\nu \<\sigma_{z_1}\cdots\sigma_{z_{2n}}\>^2_\nu &=&
 {1 \over 2^{n+2} |\eta|^2}
 \sum_{
 \stackrel{\epsilon_j=\pm1}{\sum_j \epsilon_j=0}}
 \left| \theta_\nu \!\left({\sum_j \epsilon_j z_j \over 2}\right)\right|^2
 \prod_{i<j} \left| {\theta_1(z_i-z_j)\over
                     \theta_1^\prime(0)}\right|^{\epsilon_i \epsilon_j / 2}
   \qquad \\[1mm]
&=&
 \label{def_ising_spin_correlators_full}
 {\theta_1^\prime(0)^{n/2} \over 2^{n+2} |\eta|^2}
 \sum_{
 \stackrel{\epsilon_j=\pm1}{\sum_j \epsilon_j=0}}
 \left| \theta_\nu \!\left({\sum_j \epsilon_j z_j \over 2}\right)\right|^2
 \prod_{i<j} \left| \theta_1(z_i-z_j)
             \right|^{\epsilon_i \epsilon_j / 2}
   \qquad
\end{subeqnarray}
Here we have used the complex-number notation $z = x_1+i x_2$;
$\theta'_1(0) \approx 2.8486946040$ is
the derivative of $\theta_1(z,\tau)$ with respect to $z$ evaluated at
$z=0$ and $\tau=i$;
and $\eta \approx 0.7682254223$ is the usual Dedekind function $\eta(\tau)$
evaluated at $\tau=i$.
Please note that $\theta'_1(0) = 2\pi \eta^3$
   [cf.\ (\protect\ref{eq_A.13})].
Note also that the contribution of $\{\epsilon_j\}$
to \reff{def_ising_spin_correlators_full}
is equal to that of $\{-\epsilon_j\}$,
so in the numerical evaluation of this expression
we need only take half the terms.
The expression \reff{def_ising_spin_correlators}
gives the FSS limit for the Ising-model
correlation functions at criticality:
here $z_i$ denotes the position in lattice units
divided by the lattice linear size $L$.

\bigskip

\noindent
{\bf Remark}. Although the sector $\nu=1$ does not contribute
to the partition function [since $Z_1 \sim \theta_1(0)=0$],
it does contribute to the correlation functions
[since $Z_1 \<\sigma_{z_1}\cdots\sigma_{z_{2n}}\>_1 \neq 0$].
So this sector cannot simply be discarded.
See ref.~\cite{Di_Francesco_87} for details.

\bigskip

The two correlators that are needed in the evaluation of the Binder cumulant
are
\begin{eqnarray}
\label{def_ising_2point}
Z_\nu \<\sigma_{z_1}\sigma_{z_2}\>_\nu &=&
       {|\theta'_1(0)|^{1/4} \over 2 |\eta|}
       {\left|\theta_\nu \!\left({z_1-z_2\over2}\right)\right|
        \over
       |\theta_1(z_1 - z_2)|^{1/4}} \\[4mm]
\label{def_ising_4point}
Z_\nu \<\sigma_{z_1}\sigma_{z_2} \sigma_{z_3}\sigma_{z_4} \>_\nu  &=&
  {|\theta'_1(0)|^{1/2}\over 2\sqrt{2} |\eta|}  \left\{
          \left|\theta_\nu \!\left({z_1+z_2-z_3-z_4\over2}\right)\right|^2
 \right. \nonumber \\
 & & \qquad \times
          \left| {\theta_1(z_1-z_2)\theta_1(z_3-z_4) \over
                  \theta_1(z_1-z_3)\theta_1(z_1-z_4)
                  \theta_1(z_2-z_3)\theta_1(z_2-z_4)} \right|^{1/2}
          \nonumber \\
      & & \left. \phantom{1\over1}
           + (2\leftrightarrow3) + (2\leftrightarrow4)
             \right\}^{1/2}
\end{eqnarray}
We also need the 6-point correlator to compute $V_6$. Its
exact expression can be deduced easily from the general equation
\reff{def_ising_spin_correlators_full}:
\begin{eqnarray}
& &
Z_\nu \<\sigma_{z_1}\sigma_{z_2} \sigma_{z_3}\sigma_{z_4}
        \sigma_{z_5}\sigma_{z_6} \>_\nu  =
  {|\theta'_1(0)|^{3/4}\over 4 |\eta|}  \left\{
       \left|\theta_\nu \!\left({z_1+z_2+z_3-z_4-z_5-z_6\over2}\right)\right|^2
                                        \right.
  \nonumber \\
  & & \qquad \qquad \phantom{1\over1} \times \Psi(z_1,z_2,z_3,z_4,z_5,z_6)
                    + (2\leftrightarrow4)  + (2\leftrightarrow5)
                    + (2\leftrightarrow6)
  \nonumber \\
      & & \qquad \qquad \phantom{1\over1}
           + (3\leftrightarrow4) + (3\leftrightarrow5)
           + (3\leftrightarrow6) + (2\leftrightarrow4;3\leftrightarrow5)
\nonumber \\
     & & \left. \qquad \qquad \phantom{1\over1}
           + (2\leftrightarrow4;3\leftrightarrow6)
           + (2\leftrightarrow5;3\leftrightarrow6)
           \right\}^{1/2}
  \label{G6}
\end{eqnarray}
where the function $\Psi$ is defined as
\be
\Psi(z_1,z_2,z_3,z_4,z_5,z_6) = \left(
         {\theta_1(z_{12})\theta_1(z_{13}) \theta_1(z_{23})
          \theta_1(z_{45})\theta_1(z_{46}) \theta_1(z_{56})
          \over
          \prod_{i=1,2,3;j=4,5,6}
          \theta_1(z_{ij}) } \right)^{1/2}
\ee
and we have used the shorthand notation $z_{ij} \equiv z_i -z_j$.

{}From these equations we can obtain the values of
$x^\star = \lim_{L\rightarrow\infty} \xi/L$
and $V_{2n}$ by numerical integration. In particular, 
the correlation length on a periodic lattice of size $L$ is defined to be
\be
\xi \;=\; {1 \over 2 \sin(\pi/L)} \left( {\chi \over F} - 1 \right)^{1/2}
   \;,
\label{def_xi}
\ee
where $\chi$ is the susceptibility
(i.e., the Fourier-transformed two-point correlation function at zero
momentum) and $F$ is the corresponding quantity at the smallest nonzero
momentum $(2\pi/L,0)$
[see \reff{def_msquare}/\reff{def_f}
 and \reff{def_susceptibility}/\reff{def_f_density} below for details].
This is a finite-lattice generalization of the second-moment
correlation length. Then, the universal amplitude $x^\star$ is given by
\be
x^\star  \;=\; {1 \over 2\pi} \left( {\chi \over F} - 1 \right) ^{\! 1/2}
\ee
where
\begin{eqnarray}
\chi &\sim& \int d^2z \, \<\sigma_0 \sigma_z\> \\
F    &\sim& \int d^2z \, \<\sigma_0 \sigma_z\> \, \cos (2 \pi x_1)
\end{eqnarray}
and $\int d^2z = \int_0^1 \int_0^1 dx_1 \; dx_2$. The details of this
computation are given in Appendix~\ref{app_integrals}.
We obtain
\begin{eqnarray}
   \int d^2z \, \<\sigma_0 \sigma_z\>   & = &
        1.55243 29546 5 \pm 0.00000 00000 4
     \label{chi_numerical}  \\[2mm]
   \int d^2z \, \<\sigma_0 \sigma_z\> \, \cos (2 \pi x_1)   & = &
        0.04656 74468 2 \pm 0.00000 00000 4
     \label{F_numerical}
\end{eqnarray}
As a result, we obtain $x^\star$ with 10 digits of precision:
\be
x^\star = 0.90504 88292 \pm 0.00000 00004   \;.
  \label{xstar_theor}
\ee
We repeated the computation requiring 11 digits of precision
in the integrals, and the result was the same.

The universal moment ratio $V_4$ is given by
\be
V_4 \;=\; { \int d^2z_2 \, d^2z_3 \, d^2z_4 \,
        \<\sigma_{0}\sigma_{z_2} \sigma_{z_3}\sigma_{z_4} \> \over
        \left[
        \int d^2z \, \<\sigma_0 \sigma_z\>
        \right]^2 }
   \;.
 \label{ratio_V4}
\ee
Di Francesco {\em et al.}\/ \cite{Di_Francesco_87,Di_Francesco_88}
performed the integrals in numerator and denominator
by Monte Carlo and obtained
\be
V_4 \;=\; 1.168 \pm 0.005   \;.
  \label{V4_DiFrancesco}
\ee
We have improved this value, as follows:
For the denominator of \reff{ratio_V4},
we use the very precise estimate \reff{chi_numerical}
coming from deterministic numerical integration.
For the numerator, we performed a Monte Carlo integration
using $10^9$ measurements.
Our result is
\be
\label{V4_exact}
V_4 \;=\; 1.1679 229 \pm 0.0000 047  \;,
\ee
which is compatible with \reff{V4_DiFrancesco}
but three orders of magnitude more precise.
This value also agrees closely with the estimate
of Kamieniarz and Bl\"ote \cite{Kamieniarz_93}
based on extrapolation of the exact results
(computed by transfer-matrix methods) for $L \le 17$:
\be
V_4 \;=\; 1.1679 296 \pm 0.0000 014   \;,
\label{v4_Kamieniarz}
\ee
where the error bar is of course somewhat subjective.\footnote{
   Unfortunately, Kamieniarz and Bl\"ote \cite{Kamieniarz_93}
   reported only meager details of the fits that led to this
   extraordinarily precise estimate.  
   That is a shame,
   as information on the presence or absence of particular
   correction-to-scaling terms could be of considerable
   theoretical interest.
}

More generally, the universal moment ratio $V_{2n}$ is given by
\be
V_{2n} \;=\; { \int d^2z_2 \, \cdots \, d^2z_{2n} \,
    \<\sigma_{0}\sigma_{z_2} \cdots \sigma_{z_{2n}}
    \> \over
        \left[
        \int d^2z \, \<\sigma_0 \sigma_z\>
        \right]^n }
    \;.
\ee
We have been able to compute the
(exact except for the numerical integration)
values of the ratios $V_6$, $V_8$ and $V_{10}$.
We performed the integrals in the numerator by Monte Carlo,
using $10^9$ measurements for $V_6$, $4\times 10^6$ measurements for $V_8$
and $2.5\times 10^6$ measurements for $V_{10}$.
We obtain
\begin{eqnarray}
V_6     & = & 1.4556491 \pm 0.0000072    \label{V6_exact}  \\
V_8     & = & 1.89252   \pm 0.00018      \label{V8_exact}  \\
V_{10}  & = & 2.53956   \pm 0.00034      \label{V10_exact}
\end{eqnarray}
In general, the formula for the $2n$-point function contains
$(2n)! /[ 2 (n!)^2 ]$ terms
[this takes into account the
$\{\epsilon_j\} \leftrightarrow \{-\epsilon_j\}$ symmetry],
and this grows asymptotically like $4^n$.
Thus, in computing $V_4$ (resp.\ $V_6$, $V_8$, $V_{10}$)
we had to include 3 (resp.\ 10, 35, 126) terms,
and the computation of $V_{12}$ would require handling 462 terms.
Moreover, the numerator has to be integrated over a $(4n-2)$-dimensional torus.
These facts make the high-precision numerical integration of $V_{2n}$
extremely time-consuming as soon as $n$ becomes moderately large.

Let us consider, finally, the dimensionless renormalized
four-point coupling constant $g_4$ defined by
\be
   g_4 \;=\; - \, {\bar{u}_4 \over \chi^2 \xi^d}
       \;=\; - \, {U_4 \over (\xi/L)^d}
   \;,
\ee
where $\bar{u}_4$ is the connected four-point function at zero momentum,
and more generally the dimensionless renormalized $2n$-point coupling constant
$g_{2n}$ defined by
\be
   g_{2n} \;=\;  {\chi^n \, \bar{\Gamma}_{2n} \over \xi^{(n-1)d}}
   \;,
\ee
where $\bar{\Gamma}_{2n}$ is the amputated one-particle-irreducible
$2n$-point function at zero momentum.
In the FSS limit
$L \to \infty$, $\beta \to \beta_c$ with $\xi/L$ fixed,
$g_{2n}$ becomes a nontrivial function of the FSS variable $\xi/L$,
\be
   g_{2n} \;=\; F_{g_{2n}}(\xi/L)   \;.
\ee
(There is some evidence that $F_{g_4}$ is a
{\em decreasing}\/ function of $\xi/L$.\footnote{
   Baker and Kawashima \cite{Baker_96} conjecture that
   $g_4(\beta,L)$ [resp.\ $\xi(\beta,L)$]
   is a decreasing [resp.\ increasing] function of $\beta$
   for each fixed $L < \infty$;
   these two facts, if true, would immediately imply that
   $F_{g_4}(\xi/L)$ is a decreasing function of its argument $\xi/L$.
   Numerical data for the 2D \cite{Kim-Patrascioiu_93,Baker_94}
   and three-dimensional \cite{Kim-Patrascioiu_93,Baker_96} Ising models
   clearly support the Baker--Kawashima conjecture.
})
In particular, the massive and massless scaling limits
\begin{eqnarray}
   g_{2n}^*  & = & \lim\limits_{\beta \uparrow \beta_c}
                \lim\limits_{L \to \infty}
                g_{2n}(\beta,L)   \\[1mm]
   G_{2n}^*  & = & \lim\limits_{L \to \infty}
                \lim\limits_{\beta \uparrow \beta_c}
                g_{2n}(\beta,L)
        \;=\; \lim\limits_{L \to \infty}
                g_{2n}(\beta_c,L)
\end{eqnarray}
correspond to the two extreme cases
$g_{2n}^* = F_{g_{2n}}(0)$, $G_{2n}^* = F_{g_{2n}}(x^\star)$.
The best currently available estimates for the 2D Ising model are
\begin{eqnarray}
   g_4^*  & = & \cases{14.694 \pm 0.002  & by high-temperature expansion
                                     \protect\cite{Butera_96,Pelissetto_98} \cr
                     \noalign{\vskip 2mm}
                     14.66 \pm 0.42   & by $\epsilon$-expansion
                                               \protect\cite{Pelissetto_98} \cr
                     \noalign{\vskip 2mm}
                     15.50 \pm 0.84   & by $g$-expansion
                                               \protect\cite{LeGuillou_80}  \cr
                     \noalign{\vskip 2mm}
                     14.66 \pm 0.06    & by expansion around $d=0$
                                         \protect\cite{Bender_93,Bender_95} \cr
                     \noalign{\vskip 2mm}
                     14.7 \pm 0.2      & by Monte Carlo
                                         \protect\cite{Kim_99}              \cr
                    }
      \\[4mm]
   G_4^*  & = &  2.239 \pm 0.007 \qquad\hbox{by Monte Carlo
                                         \protect\cite{Kim_99}}
      \label{Gstar4_MC_Kim}   \\[4mm]
   g_6^*  & = &  \cases{794.1 \pm 0.6   & by high-temperature expansion
                       \protect\cite{Zinn_96b,Pelissetto_98,Pelissetto_98a} \cr
                     \noalign{\vskip 2mm}
                        797 \pm 9       & by $\epsilon$-expansion
                                \protect\cite{Pelissetto_98,Pelissetto_98a} \cr
                     \noalign{\vskip 2mm}
                        792 \pm 40  & by $g$-expansion with constrained $g_4^*$
                                                  \protect\cite{Sokolov_98} \cr
                     \noalign{\vskip 2mm}
                        691 \pm 29  & by expansion around $d=0$
                                         \protect\cite{Bender_93,Bender_95} \cr
                     \noalign{\vskip 2mm}
                        850 \pm 25      & by Monte Carlo
                                         \protect\cite{Kim_99}              \cr
                       }
      \\[4mm]
   G_6^*  & = &  29.34 \pm 0.20  \qquad\hbox{by Monte Carlo
                                         \protect\cite{Kim_99}}
      \label{Gstar6_MC_Kim}   \\[4mm]
   g_8^*  & = &  \cases{(82.5 \pm 0.6) \times 10^3
                                            & by high-temperature expansion
                                \protect\cite{Pelissetto_98,Pelissetto_98a} \cr
                        \noalign{\vskip 2mm}
                        (83.8 \pm 3.2) \times 10^3  & by $\epsilon$-expansion
                                \protect\cite{Pelissetto_98,Pelissetto_98a} \cr
                        \noalign{\vskip 2mm}
                        (89 \pm 5) \times 10^3      & by Monte Carlo
                                         \protect\cite{Kim_99}              \cr
                       }
      \\[4mm]
   G_8^*  & = &  947 \pm 10      \qquad\hbox{by Monte Carlo
                                         \protect\cite{Kim_99}}
      \label{Gstar8_MC_Kim}   \\[4mm]
   g_{10}^*  & = &  \cases{(12.8 \pm 0.7) \times 10^6
                                            & by high-temperature expansion
                                \protect\cite{Pelissetto_98,Pelissetto_98a} \cr
                        \noalign{\vskip 2mm}
                           (8.0 \pm 1.4) \times 10^6
                                            & by $\epsilon$-expansion
                                \protect\cite{Pelissetto_98,Pelissetto_98a} \cr
                       }
\end{eqnarray}
Our own Monte Carlo data, reported in Section \ref{sec4.Vn},
combined with the theoretical value \reff{xstar_theor} for $x^\star$,
improve \reff{Gstar4_MC_Kim}/\reff{Gstar6_MC_Kim}/\reff{Gstar8_MC_Kim}
to $G_4^* = 2.23685 \pm 0.00016$, $G_6^* = 29.2602 \pm 0.0047$
and $G_8^* = 942.91 \pm 0.25$, respectively,
and also give values for $G_{10}^*$ through $G_{20}^*$
[see \reff{G_2n_MC_results} and the footnote following it].
{}From \reff{xstar_theor} and \reff{V4_exact}--\reff{V10_exact}
we obtain the theoretical predictions
\begin{subeqnarray}
G_4^* &=& - \, {U_4  \over  x^{\star 2}} \;=\;  2.2366587 \pm 0.0000057 \\[1mm]
G_6^* &=& - \, {U_6 - 10 U_4^2 \over x^{\star 4}}
                                          \;=\;  29.25457 \pm 0.00015 \\[1mm]
G_8^* &=& - \, {U_8 - 56 U_4 U_6 + 280 U_4^3  \over x^{\star 6}}
                                             \;=\;  942.6095  \pm 0.0072 \\[1mm]
G_{10}^* &=& - \, {U_{10} - 120 U_4 U_8 - 126 U_6^2 + 4620 U_4^2 U_6
                     - 15400 U_4^4   \over x^{\star 8}} \nonumber \\
         & &  \hspace{4.5cm} =\; 56110.24  \pm 0.56 
   \label{G_2n_exact}
\end{subeqnarray} 
The error bars in \reff{G_2n_exact} are obtained by carefully propagating
the statistical errors from the $V_{2n}$ [in which the errors are independent
except for an extremely small effect arising from the value of the
denominator \reff{chi_numerical}]
to the $U_{2n}$ (in which the errors are correlated)
and thence to the $G_{2n}^*$.

\section{Numerical Simulations} \label{sec_simul}

We consider the two-dimensional nearest-neighbor Ising model
on a $L \times L$ square lattice with periodic boundary conditions,
given by the Hamiltonian
\begin{subeqnarray}
\slabel{Hamiltonian_Ising}
{\cal H}_{\rm Ising}  &=&   -{\beta\over 2}
                             \sum_{\< ij \>} \sigma_i\sigma_j \\
\slabel{Hamiltonian_Potts}
                      &=&   -\beta \sum_{\< ij \>} \delta_{\sigma_i,\sigma_j}
                                 \,+\, \hbox{const}   \;.
\end{subeqnarray}
Note that we use throughout this paper a non-standard normalization of $\beta$,
which is motivated by considering the Ising model as a
special case of the $q$-state Potts model;
it differs by a factor of 2 from the usual Ising normalization.
In our normalization, the critical point is at
\be
  \beta_c \;=\; \log (1 + \sqrt{2}) \;\approx\; 0.881373587 \;.
\ee

\subsection{Observables to be measured}

We have performed simulations of this system using the  
Swendsen-Wang algorithm \cite{Swendsen_87,Edwards_Sokal,Sokal_Cargese}.
In particular, we have measured the following basic observables:
\begin{itemize}
   \item the energy density (i.e., the number of unsatisfied bonds)
\be
{\cal E} \;\equiv\; \sum_{\<xy\>} (1 - \delta_{\sigma_x,\sigma_y} )
\label{def_energy}
\ee
   \item the bond occupation
\be
 {\cal N} \;\equiv\; \sum_{\<xy\>} n_{xy}
\label{def_n}
\ee
   \item the nearest-neighbor connectivity
        (which is an energy-like observable \cite{Salas_Sokal_Potts3})
\be
{\cal E}' \;\equiv\; \sum_{\<xy\>} \gamma_{xy} \;,
\label{def_energyp}
\ee
where $\gamma_{xy}$ equals 1 if both ends of the bond $\<xy\>$ belong to the
same cluster, and 0 otherwise.
More generally, the connectivity $\gamma_{ij}$ can be defined
for an arbitrary pair $i,j$ of sites:
\be
  \gamma_{ij}(\{n\}) \;=\; \left\{ \begin{array}{ll}
       1 & \quad \hbox{\rm if $i$ is connected to $j$} \\
       0 & \quad \hbox{\rm if $i$ is not connected to $j$}
       \end{array} \right.
\ee
   \item the squared magnetization
\begin{subeqnarray}
{\cal M}^2  &=& \left( \sum_x \bsigma_x \right)^2        \\
            &=& {q \over q-1} \sum_{\alpha=1}^q \left(
     \sum_x \delta_{\sigma_x,\alpha} \right)^2 - {V^2 \over q-1}
 \label{def_msquare}
\end{subeqnarray}
where $\bsigma_x \equiv {\bf e}^{(\sigma_x)} \in \R^{q-1}$
is the Potts spin in the hypertetrahedral representation\footnote{
   Let $\{ {\bf e}^{(\alpha)} \} _{\alpha=1}^q$
   be unit vectors in $\R^{q-1}$ satisfying
   ${\bf e}^{(\alpha)} \cdot {\bf e}^{(\beta)} =
    (q \delta^{\alpha\beta} - 1)/(q-1)$,
   and let $\bsigma_x \equiv {\bf e}^{(\sigma_x)}$.
   For $q=2$ this means $\bsigma_x = \cos (\pi \sigma_x) = \pm 1$.
}
and $V=L^2$ is the number of lattice sites
   \item powers of the squared magnetization
\be
   {\cal M}^{2n}   \;=\;  ({\cal M}^2)^n
\ee
   \item the square of the Fourier transform of the spin variable
at the smallest allowed non-zero momentum
\begin{subeqnarray}
{\cal F}    &=&
{1 \over 2} \left( \left| \sum_x \bsigma_x \, e^{2\pi i x_1/L} \right|^2   +
                   \left| \sum_x \bsigma_x \, e^{2\pi i x_2/L} \right|^2
            \right)
     \\
  &=& {q \over q-1} \times {1 \over 2} \sum_{\alpha=1}^q \left(
   \left| \sum_x \delta_{\sigma_x,\alpha} \, e^{2\pi i x_1/L} \right|^2 +
   \left| \sum_x \delta_{\sigma_x,\alpha} \, e^{2\pi i x_2/L} \right|^2 \right)
 \label{def_f}
\end{subeqnarray}
where $(x_1,x_2)$  are the Cartesian coordinates of point $x$.
Note that ${\cal F}$ is normalized to be comparable to
its zero-momentum analogue ${\cal M}^2$.
  \item  the mean-square and mean-fourth-power size of the clusters
\begin{eqnarray}
{\cal S}_2  &=& \sum_{\cal C} \#({\cal C})^2    \label{def_s2}   \\
{\cal S}_4  &=& \sum_{\cal C} \#({\cal C})^4    \label{def_s4}
\end{eqnarray}
where the sum is over all the clusters ${\cal C}$ of activated bonds
and $\#({\cal C})$ is the number of sites in the cluster ${\cal C}$.
\end{itemize}

\noindent
{}From these observables we compute the following expectation values:
\begin{itemize}
   \item the energy density $E$ per spin
\be
E  \;=\; {1 \over V} \< {\cal E} \>
\label{def_energy_density}
\ee
   \item the specific heat
\be
C_H \;=\; {1 \over V} \var({\cal E})  \equiv {1\over V} \left[
         \< {\cal E}^2 \> - \< {\cal E} \>^2 \right]
\label{def_cv}
\ee
   \item the magnetic susceptibility
\be
  \chi \;=\; {1 \over V} \< {\cal M}^2 \>
\label{def_susceptibility}
\ee
   \item the higher magnetization cumulants
\be
   \bar{u}_{2n} \;=\; {1 \over V} \< {\cal M}^{2n} \>_{\rm conn}
\label{def_ubar_2n}
\ee
   \item the magnetization moment ratios
\be
  V_{2n} \; = \; {\< {\cal M}^{2n} \> \over \< {\cal M}^2 \>^n}
\ee
   \item  the correlation function at momentum $(2\pi/L,0)$
\be
 F \;=\; {1 \over V} \< {\cal F} \>
 \label{def_f_density}
\ee
   \item the second-moment correlation length
\be
  \xi \;=\; {1 \over 2 \sin(\pi/L)} \left( {\chi \over F} - 1 \right)^{1/2}
  \label{def_xi_sec3}
\ee
   \item the variant second-moment correlation length
\be
  \xi' \;=\; {L \over 2\pi} \left( {\chi \over F} - 1 \right)^{1/2}   \;,
  \label{def_xi'_sec3}
\ee
which differs from $\xi$ only by correction-to-scaling terms of order $L^{-2}$
\end{itemize}

\bigskip

\noindent
{\bf Remarks.}
1.  Using the Fortuin--Kasteleyn identities
\cite{Kasteleyn_69,Fortuin_Kasteleyn_72,Fortuin_72,Sokal_Cargese},
it is not difficult to show that
\begin{eqnarray}
   \< {\cal N} \>  & = &  p (B- \< {\cal E} \>)     \label{check_NE}   \\[1mm]
   \< {\cal E} \>  & = &  {q \over q-1} (B- \< {\cal E}' \>)
                                                  \label{check_EE'}   \\[1mm]
   \< {\cal M}^2 \>    & = &   \< {\cal S}_2 \>   \label{check_stwo}  \\[1mm]
   \< {\cal M}^4 \>    & = &   {q+1\over q-1} \< {\cal S}_2^2 \> -
                               {2  \over q-1} \< {\cal S}_4 \>
                                                  \label{check_sfour} 
\end{eqnarray}
where $p = 1 - e^{-\beta}$ and $B=2V$ is the number of bonds in the lattice.
%
%
% M^4  =  \sum_{x,y,z,w} (\bsigma_x . \bsigma_y) (\bsigma_z . \bsigma_w)
%
% Let's compute the conditional expectation of
%   (\bsigma_x . \bsigma_y) (\bsigma_z . \bsigma_w)
% given the bond variables {n}.
%
% If x is connected to y and z is connected to w (whether or not x,y,z,w are
%   all connected), (\bsigma_x . \bsigma_y) (\bsigma_z . \bsigma_w) = 1.
%
% If x <--> z and y <--> w but x NOT <--> y, we have
%  (\bsigma_x . \bsigma_y) (\bsigma_z . \bsigma_w) = (\bsigma_x . \bsigma_y)^2,
%  hence it's a random variable equal to 1 with probability 1/q and
%  equal to [-1/(q-1)]^2 with probability (q-1)/q, hence mean = 1/(q-1).
%
% Same if x <--> w and y <--> z but x NOT <--> y.
%
% Hence E[ (\bsigma_x . \bsigma_y) (\bsigma_z . \bsigma_w) | {n} ]
%   =  \gamma_{xy} \gamma_{zw}  +  1/(q-1) [\gamma_{xz} \gamma_{yw}
%                                 + \gamma_{xw} \gamma_{yz} - 2 \gamma_{xyzw}]
%
% and thus <M^4> = (q+1)/(q-1) <S_2^2>  -  2/(q-1) <S_4>
%
%
As a check on the correctness of our simulations,
we have tested these identities to high precision,
in the following way:
Instead of comparing directly the left and right sides of each equation,
which are strongly positively correlated in the Monte Carlo simulation,
a more sensitive test
is to define new observables corresponding to the differences
(i.e., ${\cal N} - p(B - {\cal E})$ and so forth).
Each such observable should have mean zero,
and the error bars on the sample mean
can be estimated using the standard error analysis outlined below.
The comparison to zero yields the following $\chi^2$ values:
\begin{eqnarray}
  \hbox{For \reff{check_NE}:} & \; &
     \chi^2 = 10.23 \hbox{ (14 DF, level = 75\%)}   \\
  \hbox{For \reff{check_EE'}:} & \; &
     \chi^2 = 17.00 \hbox{ (14 DF, level = 26\%)}   \\
  \hbox{For \reff{check_stwo}:} & \; &
     \chi^2 = \phantom{1}9.93  \hbox{ (14 DF, level = 77\%)}   \\
  \hbox{For \reff{check_sfour}:} & \; &
     \chi^2 = \phantom{1}9.05  \hbox{ (14 DF, level = 83\%)}
\end{eqnarray}
Here DF means the number of degrees of freedom,
and ``level'' means the confidence level of the fit
(defined at the beginning of Section \ref{sec_res_static} below).
The agreement is excellent.

2.  We also compared our data for $V_4$ for $L=4,6,8,12,16$
with the exact values computed by Kamieniarz and Bl\"ote \cite{Kamieniarz_93}
using transfer-matrix methods. We get $\chi^2 = 5.14$ (5 DF, level = 40\%),
indicating good agreement.

\bigskip
\medskip

For each observable ${\cal O}$ discussed above we have measured its
autocorrelation functions in the Swendsen-Wang dynamics,
\begin{eqnarray}
C_{\cal OO}(t)    &=& \<{\cal O}_s {\cal O}_{s+t}\> - \<{\cal O}\>^2 \\[2mm]
\rho_{\cal OO}(t) &=& {C_{\cal OO}(t) \over C_{\cal OO}(0) }
\end{eqnarray}
where the expectations are taken in equilibrium.
{}From these functions we have estimated the corresponding integrated
autocorrelation time
\begin{subeqnarray}
   \tau_{{\rm int},{\cal O}}   & = &
       {1 \over 2} \sum\limits_{t=-\infty}^\infty  \rho_{\cal OO}(t)   \\[2mm]
   & = &  {1 \over 2} \,+\, \sum\limits_{t=1}^\infty  \rho_{\cal OO}(t)
\end{subeqnarray}
by the methods of Ref.~\cite[Appendix C]{Madras_Sokal},
using a self-consistent truncation window
of width $6 \tau_{{\rm int},{\cal O}}$. This autocorrelation time is
needed to compute the correct error bar on the sample mean $\overline{\cal O}$.

\bigskip

\noindent
{\bf Remarks.} 
1. The error bar of the second-moment correlation length
is computed by considering the random variable
\be
{\cal O}' \;=\;  { {\cal M}^2 \over \mu_{{\cal M}^2} } -
                 { {\cal F}   \over \mu_{{\cal F}}   }
   \;,
\label{def_O'}
\ee
which automatically has zero mean. Then,
\be
\var(\widehat{\xi})^{1/2} \;=\;  {1 \over 4\sin(\pi/L)} {\chi \over F}
                 \left({\chi\over F} - 1\right)^{-1/2} \var({\cal O}')^{1/2}
   \;,
\ee
where $\widehat{\xi}$ denotes our Monte Carlo estimate of $\xi$.
In practice, the values of $\mu_{{\cal M}^2}$ and $\mu_{{\cal F}}$
are replaced by their corresponding sample means (which should be
computed first).

2. The error bar on the ratio $V_{2n}$ is computed in a similar fashion:
\be
\var(\widehat{V}_{2n})^{1/2} \;=\;
  { \< {\cal M}^{2n} \> \over  \< {\cal M}^2 \>^n }
                 \var({\cal O}''_{2n})^{1/2}  \;,
\ee
where $\widehat{V}_{2n}$ denotes our Monte Carlo estimate of $V_{2n}$,
and the observable ${\cal O}''_{2n}$ is defined as
\be
{\cal O}''_{2n} \;=\; { {\cal M}^{2n} \over \mu_{{\cal M}^{2n}} } -
                 n{ {\cal M}^2    \over \mu_{{\cal M}^{2}}  }
                  + n-1
\ee
and has mean zero.
Again, the mean values $\mu_{{\cal M}^{2n}}$ and $\mu_{{\cal M}^2}$
are replaced in practice by their sample means.

3.  As a further check on the correctness of our simulations,
we have computed both sides of the identity
\be
\rho_{\cal NN}(1) \;=\; 1 - { (1-p)(2 - E) \over p C_H + (1-p)(2 - E) }
\ee
proven in \cite[equation 7]{Li_Sokal}
(see also \cite{Salas_Sokal_Potts3}).\footnote{
   Please note that \protect\cite{Li_Sokal} used a definition of energy
   that is slightly different from the one used here:
   $E({\rm Ref.~\protect\cite{Li_Sokal}}) = (1/V)\<\sum_{\<xy\>}
    \delta_{\sigma_x,\sigma_y}\> = 2 - E$.
}
This is a highly nontrivial test, as it relates static quantities
(energy and specific heat) to a dynamic quantity (autocorrelation function
of the bond occupation at time lag 1). We have also checked with great
accuracy the identities \cite{Salas_Sokal_Potts3}
\begin{eqnarray}
C_{\cal EE}(t)      &=& {1 \over p^2} C_{\cal NN}(t+1) \\
\rho_{\cal EE}(t)   &=& {\rho_{\cal NN}(t+1) \over \rho_{\cal NN}(1)} \\
C_{\cal E'E'}(t)    &=& \left({q \over q-1}\right)^2 C_{\cal EE}(t+1) \\
\rho_{\cal E'E'}(t) &=& {\rho_{\cal EE}(t+1) \over \rho_{\cal EE}(1)}
\end{eqnarray}

\subsection{Summary of the simulations}

We have run our Monte Carlo
program on lattices with $L$ ranging from 4 to 512
(see Table~\ref{table_statics}).
In all cases the initial configuration was random,
and for $L\leq 64$ (resp.\ $L\geq 96$)
we discarded the first $5 \times 10^4$ (resp.\ $10^5$) iterations
to allow the system to reach equilibrium; this discard interval is
in all cases greater than $10^4 \, \tau_{{\rm int},{\cal E}}$.\footnote{
   Such a discard interval might seem to be much larger than necessary:
   $10^2 \tau_{\rm int}$ would usually be more than enough.
   However, there is always the danger that the longest autocorrelation time
   in the system may be much larger than the longest autocorrelation time
   that one has {\em measured}\/, because one has failed to measure
   an observable having sufficiently strong overlap with the slowest mode.
   As an undoubtedly overly conservative precaution against the possible
   (but unlikely) existence of such a (vastly) slower mode,
   we decided to discard up to 2\% of the entire run.
   This discard amounts to reducing the accuracy on our final estimates
   by a mere 1\%.
}
The total number of iterations ranges from
$2.15 \times 10^6$ ($L=4$) to $8.2 \times 10^6$ ($L=512$),
and is selected to be approximately $10^6 \, \tau_{{\rm int},{\cal E}}$.
These statistics allow us to obtain a high accuracy in our
estimates of the static and dynamic quantities
(error $\ltapprox$ 0.17\% and $\ltapprox$ 0.51\%, respectively).
The static data are displayed in
Table~\ref{table_statics} ($\chi,F,\xi$) 
and Table~\ref{table_ratios} (the ratios $V_{2n}$).
The dynamic data will be reported elsewhere.

The CPU time required by our program is approximately
6.3 $L^2$ $\mu$s per iteration on a Linux Pentium machine
running at 166 MHz.
The total CPU time used in
the project was approximately 7.5 months on this machine.

We have improved the precision of our analysis of the correlation length 
$\xi$ by supplementing our own Monte Carlo data with comparable data from 
Ballesteros {\em et al.}\/ \cite{complutense}.
They performed single-cluster \cite{Wolff_89}
simulations of the 2D site-diluted Ising model at various
concentrations ${\sf p}$.
Their data for ${\sf p}=1$ (i.e., the usual Ising model)
correspond to anywhere from $4 \times 10^5$ to $7 \times 10^5$
statistically independent measurements
at each lattice size from $L=12$ to $L=512$
(see Table~\ref{table_xi_JJ}). The statistical independence of two
consecutive measurements was achieved by allowing 100 single-cluster moves
between them.
Their error bars are slightly larger than ours. As a matter of fact, their
error bars $\sigma'(\xi)$ and our error bars $\sigma(\xi)$ satisfy
approximately the relation
\be
{\sigma(\xi) \over \sigma'(\xi)}
   \;=\; \sqrt{2 \tau_{{\rm int},{\cal O}'} N' \over N} \;\,,
\ee
where $N$ (resp. $N'$) is the number of measurements
of our (resp.\ their) work,
and ${\cal O}'$ is the observable \reff{def_O'}
we used to compute the correct correlation-length error bar.
This supports the belief that their measurements are indeed
essentially independent and that their error bars are correctly computed.
Comparison of their raw data to ours at the eleven overlapping $L$ values
yields $\chi^2 =  10.28$ (11 DF, level = 51\%).
The two data sets are therefore compatible.
The corresponding merged data are shown in Table~\ref{table_xi_merged}.

\section{Data Analysis} \label{sec_res_static}

For each quantity ${\cal O}$, we carry out a variety of fits
using the standard weighted least-squares method.
As a precaution against corrections to scaling,
we impose a lower cutoff $L \ge L_{min}$
on the data points admitted in the fit,
and we study systematically the effects of varying $L_{min}$ on the
estimated parameters and on the $\chi^2$ value.
In general, our preferred fit corresponds to the smallest $L_{min}$
for which the goodness of fit is reasonable
(e.g., the confidence level\footnote{
   ``Confidence level'' is the probability that $\chi^2$ would
   exceed the observed value, assuming that the underlying statistical
   model is correct.  An unusually low confidence level
   (e.g., less than 5\%) thus suggests that the underlying statistical model
   is {\em incorrect}\/ --- the most likely cause of which would be
   corrections to scaling.
}
is $\gtapprox$ 10--20\%)
and for which subsequent increases in $L_{min}$ do not cause the
$\chi^2$ to drop vastly more than one unit per degree of freedom (DF).

\subsection{Corrections to scaling} \label{sec4.fss}

In the data analysis we should take into account the effect of corrections 
to scaling in order to get reliable estimates of the physical quantities.
In particular, the value at criticality of any observable $O(L)$  
is typically given for large $L$ by
\be
O(L) = A L^{p_O}  \left( 1  + A' L^{-\Delta} + \cdots \right)
\ee
where $p_{O}$ is the critical exponent associated to the observable $O$,  
$\Delta$ is the leading correction-to-scaling exponent, and the dots
indicate higher-order corrections.

In finite-size-scaling (FSS) theory \cite{Privman}
for systems with periodic boundary conditions,
three simplifying assumptions have frequently been made:
\begin{itemize}
\item[(a)] The regular part of the free energy, $f_{\rm reg}$,
           is independent of $L$ \cite{Privman}
           (except possibly for terms that are exponentially small in $L$).
\item[(b)] The relations connecting the nonlinear scaling fields
           $g_t$ and $g_h$ to the conventional thermodynamic parameters
           $t \equiv \beta_c-\beta$ and $h$
           are independent of $L$ \cite{Guo_87}.
\item[(c)] The scaling field $g_L$ associated to the lattice size equals 
           $L^{-1}$ exactly, with no corrections $L^{-2}$,
           $L^{-3}$, \ldots\  \cite{Privman}.
\end{itemize}
Moreover, in the nearest-neighbor spin-1/2 2D Ising model,
it has further been assumed that
\begin{itemize}
\item[(d)]  There are no irrelevant operators \cite{Aharony_83,Gartenhaus_88}.
\end{itemize}
This latter assumption has been confirmed numerically
(in the infinite-volume theory) through order $t^3$,
at least as regards the bulk behavior of the susceptibility
\cite{Gartenhaus_88}. 
However, both numerical \cite{Nickel_99a,Nickel_99b}
and theoretical \cite{Pelissetto_private} evidence has recently emerged
suggesting that irrelevant operators {\em do}\/ contribute
to the susceptibility at order $t^4$.

The absence of irrelevant operators implies that the corrections to scaling 
in this model are due to the smooth but in general nonlinear connection between 
the conventional thermodynamic parameters $t$ and $h$
and the renormalization-group nonlinear scaling fields
\cite{Wegner_72,Wegner_D+G_vol6,Aharony_83,Gartenhaus_88}.  
Starting from the FSS Ansatz for the Ising-model free energy and using 
the above assumptions, it is possible to obtain a FSS expression for the 
usual observables at criticality as functions of the lattice size $L$ 
\cite{Salas_Sokal_FSS_paper2}. In particular, the leading correction term
in the expansion of the susceptibility is the $L$-independent term
coming from the regular part of the free energy.
This implies that for this observable
\be
   \Delta  \;=\; {7\over 4}   \;\,.
\label{Delta_prediction}
\ee
The same result is plausible for the observable $F$ defined in
\reff{def_f_density};  thus, we expect $\Delta=7/4$
for the second-moment correlation lengths $\xi$ and $\xi'$
and the corresponding amplitude $x^\star$.
The expansion for the magnetization cumulant $\bar{u}_{2n}$
gives an exponent $\Delta = 1 + \gamma/\nu = 11/4$
(perhaps with a multiplicative logarithmic correction).
Thus, we expect that the ratios $V_{2n}$ 
also have a correction-to-scaling exponent given by \reff{Delta_prediction}
[due to the power of the susceptibility appearing in its definition 
\reff{def_ratios}].  
For a more detailed theoretical and numerical analysis of
the corrections to scaling in this model, see \cite{Salas_Sokal_FSS_paper2}.

\subsection{Second-moment correlation length}   \label{sec4.xi}

The second-moment correlation length $\xi$ and its variant $\xi'$
[cf.\ \reff{def_xi_sec3}/\reff{def_xi'_sec3}]
are expected to behave as
\be
  \left\{  {\xi \atop \xi'}  \right\}
    \;=\;  L^p \left[ x^\star + A L^{-\Delta} + \cdots \right]
 \label{ansatz_xi}
\ee
with $p=1$. We can estimate $p$ by ignoring correction-to-scaling terms and
performing a simple power-law fit.  We get
\begin{eqnarray}
  \hbox{For $\xi$:}  &  &
      p = 0.99974 \pm 0.00036 \qquad
      \hbox{ ($L_{min} = 32$, $\chi^2=1.48$, 6 DF, level = 96\%)}
     \nonumber \\   \\
  \hbox{For $\xi'$:}  &  &
      p = 1.00018 \pm 0.00036 \qquad
      \hbox{ ($L_{min} = 32$, $\chi^2=1.37$, 6 DF, level = 97\%)}
     \nonumber \\
\end{eqnarray}
The agreement with the theoretical prediction is excellent.

The value of the constant $x^\star$ can be estimated most simply
by fitting the ratio $\xi/L$ or $\xi'/L$
to a constant, ignoring corrections to scaling.
We get
\begin{eqnarray}
  \hbox{For $\xi$:}  &  &
      x^\star = 0.90577 \pm 0.00028   \qquad
      \hbox{ ($L_{min} = 32$, $\chi^2=2.02$, 7 DF, level = 96\%)}
    \nonumber \\ \label{xstar_xi_simple}   \\
  \hbox{For $\xi'$:}  &  &
      x^\star = 0.90557 \pm 0.00026   \qquad
      \hbox{ ($L_{min} = 16$, $\chi^2=2.73$, 9 DF, level = 97\%)}
    \nonumber \\ \label{xstar_xi'_simple}
\end{eqnarray}
The estimate based on $\xi$ lies 2.6 standard deviations away from
the value $x^\star \approx 0.90505$ predicted by CFT
[cf.\ \reff{xstar_theor}].
The estimate based on $\xi'$ is slightly better:
it lies two standard deviations away from the theoretical prediction,
and works also for smaller $L_{min}$.
Indeed, the corrections to scaling in $\xi'/L$ are negligible
(compared to our statistical error) already for $L \ge 16$
(see the last column of Table~\ref{table_xi_merged}).\footnote{
   Table~\ref{table_xi_merged} is based on merging our data
   with that of Ballesteros {\em et al.}\ \cite{complutense};
   but virtually identical results are obtained using our data alone.
}
This fact makes it almost hopeless to study corrections to FSS
in $\xi'$.

If we fit $\xi$ to \reff{ansatz_xi},
keeping the first correction-to-scaling term and trying to
estimate simultaneously the three parameters $x^\star$, $A$ and $\Delta$,
a good fit is obtained for $L_{min} = 8$:
\begin{subeqnarray}
x^\star  &=& 0.90546 \pm 0.00033  \slabel{fit_xi_deltavar_a} \\
A        &=& 0.75    \pm 0.30 \\
\Delta   &=& 1.76    \pm 0.18     \slabel{fit_xi_deltavar_Delta}
\label{fit_xi_deltavar}
\end{subeqnarray}
with $\chi^2=2.97$ (9 DF, level = 97\%).
The value of $x^\star$ is again two standard deviations away from
the theoretical prediction \reff{xstar_theor}.
The estimate of $\Delta$ is very close to $7/4$, and is only 1.4 standard
deviations away from $2$; 
but perhaps this estimate ought not be taken too seriously,
as the correction-to-scaling amplitude is only 2.5 standard deviations
away from zero (a deviation that is, moreover, comparable to the
discrepancy in $x^\star$).
The analogous fit for $\xi'$ is even more hopeless
(the amplitude $A$ is compatible with zero within 0.7 standard deviations),
so we omit the details.
This correction-to-scaling exponent $\Delta \approx 2$
can be understood as arising simply from the ratio
$\xi/\xi' \equiv [(L/\pi) \sin (\pi/L)]^{-1} =
 1 + (\pi^2/6) L^{-2} + \ldots\;$. Indeed, if we fit the data to the
Ansatz $\xi/L = x^\star + A L^{-2}$ we get for $L_{min}=8$:
\begin{subeqnarray}
x^\star  &=& 0.90569 \pm 0.00027 \\
A        &=& 1.269 \pm 0.064
  \label{xstar_xi_Lminus2}
\end{subeqnarray}
with $\chi^2=4.56$ (10 DF, level = 92\%). Then,
$A/x^\star \approx 1.40$, which is not far from $\pi^2/6 \approx 1.64$.

We can improve the precision of our numerical estimates by
using the merged data of Table~\ref{table_xi_merged}
(our data plus that of Ballesteros {\em et al.}\ \cite{complutense}).
The simple power-law fit yields
\begin{eqnarray}
  \hbox{For $\xi$:}  &  &
      p = 1.00047 \pm 0.00033 \qquad
      \hbox{ ($L_{min} = 48$, $\chi^2=0.86$, 5 DF, level = 97\%)}
     \nonumber \\   \\
  \hbox{For $\xi'$:}  &  &
      p = 1.00074 \pm 0.00033 \qquad
      \hbox{ ($L_{min} = 48$, $\chi^2=0.82$, 5 DF, level = 98\%)} \nonumber \\
\end{eqnarray}
The fits to a constant give
\begin{eqnarray}
  \hbox{For $\xi$:}  &  &
      x^\star = 0.90565 \pm 0.00023 \qquad
      \hbox{ ($L_{min} = 48$, $\chi^2=2.92$, 6 DF, level = 82\%)}
     \nonumber \\   \\
  \hbox{For $\xi'$:}  &  &
      x^\star = 0.90555 \pm 0.00020 \qquad
      \hbox{ ($L_{min} = 16$, $\chi^2=6.99$, 9 DF, level = 64\%)} \nonumber \\
\end{eqnarray}
The three-parameter fit $\xi/L = x^\star + A L^{-\Delta}$
is good for $L_{min} = 8$:
\begin{subeqnarray}
x^\star &=& 0.90552 \pm 0.00026 \slabel{fit_xi_merged_deltavar_xstar} \\
A       &=& 0.86    \pm 0.29    \\
\Delta  &=& 1.82    \pm 0.15
\label{fit_xi_merged_deltavar}
\end{subeqnarray}
with $\chi^2 = 7.57$ (9 DF, level = 58\%).
The analogous fit with $\xi'$
yields a correction-to-scaling amplitude compatible with zero within errors.

In conclusion, one can extract accurate estimates of the critical exponent $p$
and the amplitude $x^\star$ using our Monte Carlo data;
the results agree with the theoretical prediction \reff{xstar_theor}
within two standard deviations.
However, it is very difficult to estimate from our numerical data
the correction-to-scaling exponent (or the corresponding amplitude).
Indeed, for $\xi'$ the corrections to scaling are negligible
(compared to our statistical error) for $L \ge 16$.

\subsection{Magnetization moment ratios} \label{sec4.Vn}

If we study the magnetization distribution $\rho({\cal M})$
as $L \to \infty$ at fixed $\beta$,
we expect three distinct behaviors depending on the value of $\beta$:
\begin{itemize}
\item[(a)]
At $\beta < \beta_c$, we are in the high-temperature regime, where
correlations decay exponentially.
A variant of the central limit theorem \cite{Newman_private}
guarantees that the finite-$L$ distributions will converge,
after rescaling by the factor $\sqrt{V\chi}$,
to a Gaussian distribution of mean zero and unit variance.\footnote{
   Previously this had been proven for finite subvolumes
   of an infinite system \cite{Newman_80,Newman_83},
   by a technique using the FKG inequalities.
   It has recently been proven by Newman \cite{Newman_private}
   also for finite systems with periodic or free boundary conditions,
   by a different (but simple and elegant) method using the
   GKS, GHS and Simon-Lieb inequalities.
   We thank Professor Newman for communicating to us
   this unpublished result, which we hope he will someday publish.
}

\item[(b)]
At $\beta>\beta_c$ we are in the low-temperature regime,
and the finite-$L$ distributions should converge,
after rescaling by the factor $V M_0$
(where $M_0$ is the spontaneous magnetization),
to the sum of two delta functions.
There are Gaussian fluctuations around these two delta functions,
but their width is much smaller, namely
$\sqrt{V \chi_0}$, where $\chi_0$ is the susceptibility in a pure phase.

\item[(c)]
At $\beta=\beta_c$ [or more generally, at fixed value of the
FSS variable $L^{1/\nu}(\beta-\beta_c)$],
the finite-$L$ distributions will converge, after
rescaling by the factor $\sqrt{V \chi}$, to some non-Gaussian
distribution characteristic of the critical Ising model in a finite box.
This distribution is not, to our knowledge,  known exactly.
\end{itemize}

\begin{figure}
\epsfxsize=400pt\epsffile{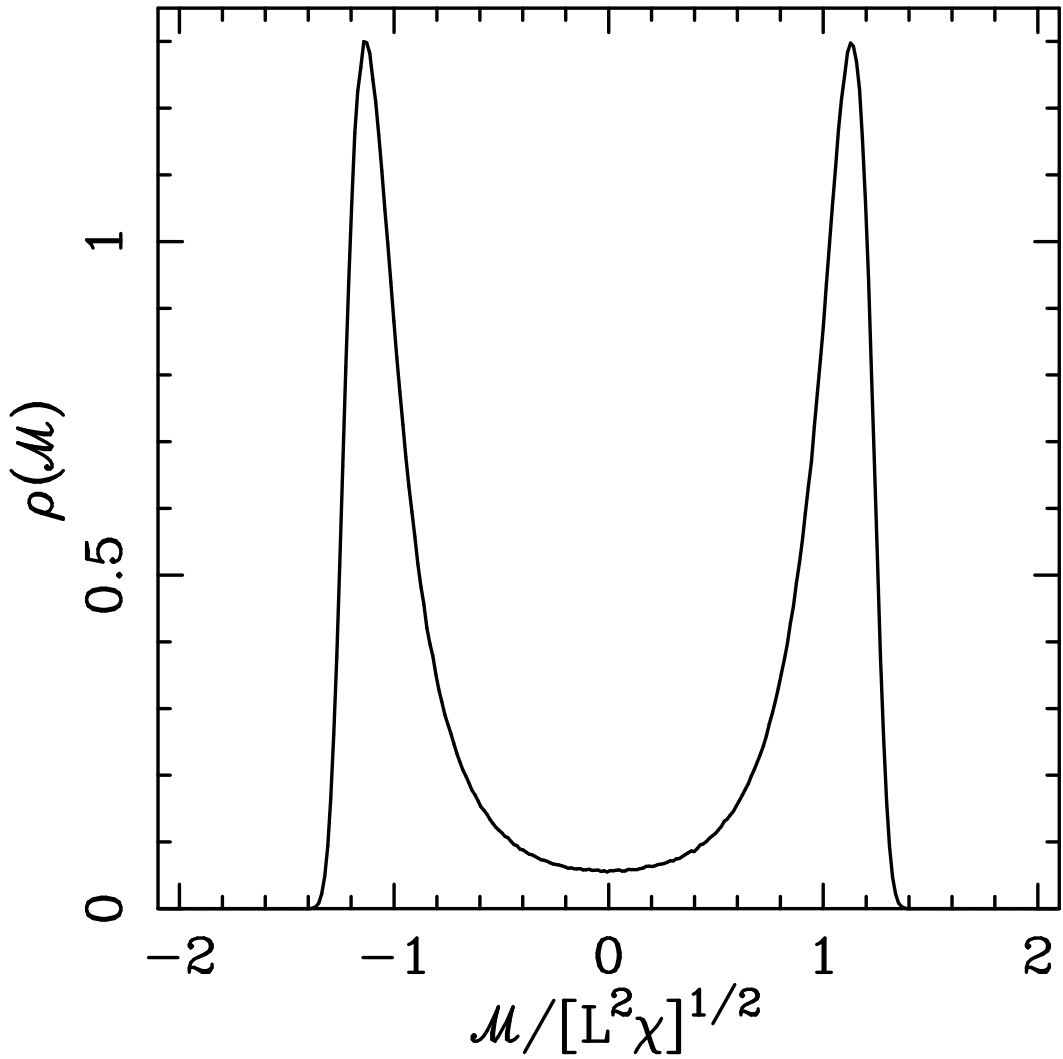}
  \caption{Magnetization histogram of the 2D Ising model at
           $\beta=\beta_c$ for $L=256$.
           The histogram is normalized such that the area enclosed is
           equal to unity.
  }
\label{figure_histo_mag_256}
\end{figure}

We have computed the magnetization histograms at $\beta=\beta_c$ for
$L=4,\ldots,512$. The sequence of histograms is expected to converge
to a limiting distribution when we normalize the magnetization
by $\sqrt{V \chi}$ and normalize the height of the bins so that
the area enclosed by the histogram is 1.
For $L\gtapprox64$ the histograms converge well to a limiting histogram
(Figure~\ref{figure_histo_mag_256}). For $L\ltapprox48$ small
corrections to scaling are observed: the peaks of the histogram
are slightly taller than in the limiting histogram
The limiting distribution is symmetric and very strongly two-peaked
(with maxima at ${\cal M}/\sqrt{V \chi} \approx \pm 1.11$);
clearly the 2D Ising model at criticality in a finite symmetric torus
is very far from Gaussian
(e.g., we will find $V_4$ much closer to 1 than to 3).

In order to characterize quantitatively this limiting distribution,
we have measured its moments $\<{\cal M}^{2n}\>$ for $n=1,\ldots,10$
and have computed the corresponding ratios
$V_{2n} \equiv \<{\cal M}^{2n}\> / \<{\cal M}^2\>^n$.
We expect a behavior
\be
  V_{2n}  \;=\;
  V_{2n}^\infty + B_{2n} L^{-\Delta} + \cdots
     \;.
  \label{ansatz_V2n}
\ee
For each $n$, we have fitted our numerical data (Table~\ref{table_ratios})
in two ways:
a one-parameter fit to a constant $V_{2n} = V_{2n}^\infty$
(fits marked with a C on the second column of Table~\ref{table_fit_ratios})
and a three-parameter fit to $V_{2n} = V_{2n}^\infty + B_{2n} L^{-\Delta}$
(fits marked P in Table~\ref{table_fit_ratios}).

As expected, the estimates of $V_{2n}^\infty$ lie in-between the
values associated to a Gaussian distribution \reff{ratios_Gaussian} and
those associated to a two-delta-function distribution \reff{ratios_2deltas}.
However, they are much closer to the latter values,
reflecting the strongly two-peaked shape of the magnetization distribution.

The fits to a constant are excellent for $L_{min} \gtapprox 32$--64;
for $2n = 4,6,8,10$ the estimates of $V_{2n}^\infty$
agree with the theoretical predictions within
about 2.5 standard deviations.
The three-parameter fits are excellent already for $L_{min}=8$:
the correction-to-scaling amplitude $B_{2n}$ grows in magnitude with $n$,
while the values of the correction-to-scaling exponent $\Delta$
are quite stable and are consistent with the theoretical prediction
$\Delta = \gamma/\nu = 7/4$ [cf.\ \reff{Delta_prediction}]
within two standard deviations.

Let us now look more closely at $V_4$.
With the three-parameter fit, we obtain for $L_{min}=8$:
\begin{subeqnarray}
V_4^\infty   &=& 1.16777 \pm 0.00013 \slabel{fit_v4_deltavar_v4} \\
\Delta       &=&    2.01 \pm 0.22    \\
B_4          &=&   -0.48 \pm 0.23
\label{fit_v4_deltavar}
\end{subeqnarray}
with $\chi^2=1.53$ (9 DF, level = 100\%).
The estimate of $V_4^\infty$ is about one standard deviation away from
the theoretical prediction $V_4^\infty \approx 1.16792$
[cf.\ \reff{V4_exact}].
The estimate of $\Delta$ is reasonably close to $\Delta=7/4$;
but since the estimated correction amplitude $B_4$ is only
2 standard deviations away from zero, this estimate
of $\Delta$ should perhaps not be taken too seriously.

Similarly, the estimates
\begin{eqnarray}
   V_6^\infty     & = &  1.45517 \pm 0.00037   \\
   V_8^\infty     & = &  1.89163 \pm 0.00079   \\
   V_{10}^\infty  & = &  2.5377 \pm 0.0015
\end{eqnarray}
from the three-parameter fit
are compatible with the predicted exact values
\reff{V6_exact}/\newline\reff{V8_exact}/\reff{V10_exact}
within 1.5, 1.3 and 1.2 standard deviations, respectively.
The estimates of the exponent $\Delta$ ($1.90 \pm 0.16$, $1.83 \pm 0.13$ and
$1.78 \pm 0.11$, respectively) are compatible with 7/4.

The values of the first nine dimensionless renormalized $2n$-point 
coupling constants at criticality on a symmetric torus
can be obtained from the results contained in Table~\ref{table_fit_ratios}: 
\begin{subeqnarray} 
G_{4\phantom{0}}^* &=& 2.23685 \pm 0.00016 \\[1mm]
G_{6\phantom{0}}^* &=& 29.2602 \pm 0.0047 \\[1mm]
G_{8\phantom{0}}^* &=& 942.91  \pm 0.25   \\[1mm]
G_{10}^* &=& (5.6135 \pm 0.0021)\times 10^4 \\[1mm]
G_{12}^* &=& (5.3281 \pm 0.0026)\times 10^6 \\[1mm]
G_{14}^* &=& (7.3681 \pm 0.0046)\times 10^8 \\[1mm]
G_{16}^* &=& (1.3969 \pm 0.0010)\times 10^{11} \\[1mm]
G_{18}^* &=& (3.4746 \pm 0.0031)\times 10^{13} \\[1mm]
G_{20}^* &=& (1.0969 \pm 0.0011)\times 10^{16} 
  \label{G_2n_MC_results}
\end{subeqnarray}
The central values in \reff{G_2n_MC_results}, which are intended as our
``best estimates'', are computed using the {\em theoretical}\/
value \reff{xstar_theor} for $x^\star$.\footnote{
   A ``pure'' Monte Carlo value, using the estimate 
   \reff{fit_xi_merged_deltavar_xstar}
   for $x^\star$ in place of \reff{xstar_theor}, can be obtained by
   dividing the central value in \reff{G_2n_MC_results} by $1.005206^{2n-2}$
   and adding $(2n-2) \times 0.000287$ to the fractional error bar.
   Note that the {\em dominant}\/ contribution to the error bar on $G_{2n}^*$
   would then come from the uncertainty on $x^\star$:
   for example, we would have $G_4^* = 2.2345 \pm 0.0014$,
   $G_6^* = 29.199 \pm 0.038$, $G_8^* = 939.97 \pm 1.87$, etc.\ 
   in place of \reff{G_2n_MC_results}.
}
The error bars quoted in \reff{G_2n_MC_results} are upper bounds 
computed using the triangle 
inequality, since we did not bother to compute the covariances among
our Monte Carlo estimates of $V_{2n}^\infty$.
Of course, for $G_4^*, G_6^*, G_8^*, G_{10}^*$
the estimates (\ref{G_2n_MC_results}a--d)
are supplanted by the much more precise theoretical values
(\ref{G_2n_exact}a--d).

In summary, we have been able to estimate the limiting values
$V_{2n}^\infty$ with great accuracy;
and in the cases where the exact values are known,
our numerical estimates agree with the theoretical predictions
within less than two standard deviations.
Our numerical estimates for $\Delta$ are compatible
(within less than two standard deviations) with $\Delta=7/4$.

\section{Conclusions} \label{sec_conclusions}

We have computed, using results from conformal field theory (CFT),
the exact (except for numerical integration) values of
five universal amplitude ratios characterizing the
2D Ising model at criticality on a symmetric torus:
the correlation-length ratio $x^\star$
and the magnetization moment ratios $V_4$, $V_6$, $V_8$ and $V_{10}$.
All except for $V_4$ are new, and we have improved
previous CFT determinations of $V_4$ by three orders of magnitude
(reaching precision similar to that obtained by transfer-matrix approaches).
As a corollary, we have computed the exact values
$G_4^*$, $G_6^*$, $G_8^*$ and $G_{10}^*$
of the first four dimensionless renormalized $2n$-point coupling constants 
at criticality on a symmetric torus.

We have checked all these theoretical predictions by means of a high-precision
Monte Carlo simulation. Using finite-size-scaling (FSS) techniques, we have
tried to determine the leading term as well as the correction-to-scaling
terms.
We confirm to high precision the theoretically predicted
universal amplitude ratios $x^\star$, $V_4$, $V_6$, $V_8$ and $V_{10}$
(error bars $\ltapprox 0.06\%$).

The determination of the leading correction-to-scaling exponent $\Delta$
has proved to be difficult.
For the modified correlation length $\xi'$,
the corrections to FSS
are so weak that they are essentially invisible for $L \ge 16$;
and no reliable conclusions can be obtained from our data for $L=4,6,8,12$.
For the standard correlation length $\xi$,
the leading correction to scaling might be $\Delta=7/4$,
or it might be $\Delta=2$ arising from
$\xi/\xi' \equiv [(L/\pi) \sin (\pi/L)]^{-1} =
 1 + (\pi^2/6) L^{-2} + \ldots\;$.
For the magnetization moment ratios $V_{2n}$ we obtain stable results 
compatible with $\Delta=7/4$ within two standard deviations, in agreement with
the theoretical prediction \reff{Delta_prediction}.

It would be interesting to extend the analytic computation of
$x^\star$ to other two-dimensional models,
in particular those that can be mapped onto Gaussian models
via height representations
(see e.g.\ \cite{Nijs_82,Burton_Henley_97,Salas_Sokal_swaf3}).
This work is currently in progress \cite{CPSS_in_prog}.

\appendix

\section{Computation of spin-correlator integrals} \label{app_integrals}

The computation of $x^\star = \lim\limits_{L\rightarrow\infty} \xi/L$
involves computing numerically the integrals
\begin{eqnarray}
I_1 &=& \int d^2z \, {\sum\limits_{\nu=1}^4 |\theta_\nu (z/2)|
                      \over
                      |\theta_1(z)|^{1/4} }     \\[2mm]
I_2 &=& \int d^2z \, {\sum\limits_{\nu=1}^4 |\theta_\nu (z/2)|
                      \over
                      |\theta_1(z)|^{1/4} }
                     \, \cos (2 \pi x_1)
\end{eqnarray}
where $z = x_1 + ix_2$ and
$\int d^2z = \int_0^1 \int_0^1 dx_1 \, dx_2$.

Let us consider here $I_1$, as $I_2$ can be done in a similar fashion.
Using the symmetry properties of the $\theta$-functions
and their absolute values (see Appendix~\ref{app_theta}),
we reduce the integral to
\be
  I_1  \;=\;  4 \int\limits_0^{1/2} \int\limits_0^{1/2}  dx_1 \, dx_2 \,
                     {\sum\limits_{\nu=1}^4 |\theta_\nu (z/2)|
                      \over
                      |\theta_1(z)|^{1/4} }
   \;\,.
\ee
The integrand contains two pieces:
One (coming from $\nu=1$) is finite at $z=0$ and its
integral can be performed safely by standard deterministic
numerical-integration techniques
(e.g. {\sc Mathematica}'s {\tt NIntegrate}), yielding
\be
  I_{1,1}  \;\equiv\;
4 \int\limits_0^{1/2} \int\limits_0^{1/2}  dx_1 \, dx_2 \,
     {|\theta_1 (z/2)|   \over   |\theta_1(z)|^{1/4} }
   \;=\;  0.52348 26517 \pm 0.00000 00001
\ee
The other piece (coming from $\nu=2,3,4$) diverges at $z=0$ like
$|\theta_1(z)|^{-1/4} \sim |z|^{-1/4}$.
This singularity makes numerical integration a bit tricky.
Since $\theta'_1(0) = 2\pi \eta^3$ [see \reff{eq_A.13}],
the simple function
\be
H(z) \;=\;  4 \, {\sum\limits_{\nu=2}^4 \left|\theta_\nu(0)\right|
                  \over
                  |2 \pi \eta^3 z|^{1/4}}
\ee
has exactly the same divergent behavior at $z=0$.
The integral of this function is given by
\begin{eqnarray}
4 \int\limits_0^{1/2} \int\limits_0^{1/2}  dx_1 \, dx_2 \, H(z) &=&
4 \, {\sum_{\nu=2}^4 \left|\theta_\nu(0)\right| \over (2 \pi \eta^3)^{1/4}}
\int_0^{1/2} \int_0^{1/2}  dx_1 \; dx_2 {1 \over (x_1^2 + x_2^2)^{1/8}}
\nonumber \\[1mm]
 &=&
8 \, {\sum_{\nu=2}^4 \left|\theta_\nu(0)\right| \over (2 \pi \eta^3)^{1/4}}
\int_0^{\pi/4} d\psi \int_0^{1/(2\cos\psi)}  dr \; r^{3/4}
\nonumber \\[1mm]
 & = & {8 \; 2^{1/4}\over 7}
{\sum_{\nu=2}^4 \left|\theta_\nu(0)\right| \over (2 \pi \eta^3)^{1/4}}
\int_0^{\pi/4} (\cos\psi)^{-7/4} \, d\psi
\nonumber \\[1mm]
&\approx& 2.95015 47241 9465
\end{eqnarray}
Though we were unable to perform exactly the final angular integral,
the integrand $\cos^{-7/4}\psi$ is regular on the interval $[0,\pi/4]$
and so the integral can be performed by
standard numerical-integration techniques.

Finally, we have to integrate the function
\be
4 \, {\sum\limits_{\nu=2}^4 |\theta_\nu(z/2)|
      \over
      |\theta_1(z)|^{1/4}}  \;-\; H(z)
   \;.
 \label{eqB.7}
\ee
This function does not diverge at $z=0$
(or at any other point in the integration domain),
so its integral can again be performed using standard techniques.
This last integral is $0.00 79738 83019 \pm 0.00 00000 00001$,
so the final result is
\be
I_1 \;=\; 3.48161 12589 \pm 0.00000 00001   \;.
\ee

The second integral $I_2$ can be performed in the same way
[and using the same auxiliary function $H(z)$]. The final result is
\be
I_2 \;=\; 0.10443 59092 \pm 0.00000 00001   \;.
\ee

\section{Theta Functions}   \label{app_theta}

We use the following definitions for the Jacobi $\theta$-functions
\cite{Itzykson,Gradshteyn}:
\begin{subeqnarray}
\theta_1(z,\tau) &\equiv& -i \sum_{n=-\infty}^{\infty} (-1)^n y^{n+{1\over2}}
                          q^{{1\over2}\left(n+{1\over2}\right)^2} \\
                 &=&      2 \sum_{n=0}^{\infty}
                          (-1)^n q^{{1\over2}\left(n+{1\over2}\right)^2}
                          \sin\left(2\pi \left(n+{1\over2}\right)z\right)
\end{subeqnarray}
\begin{subeqnarray}
\theta_2(z,\tau) &\equiv& \sum_{n=-\infty}^{\infty} y^{n+{1\over2}}
                           q^{{1\over2}\left(n+{1\over2}\right)^2} \\
                  &=&      2 \sum_{n=0}^{\infty}
                           q^{{1\over2}\left(n+{1\over2}\right)^2}
                          \cos\left(2\pi \left(n+{1\over2}\right)z\right)
                          \hspace{1cm}  %% TO LINE UP EQUAL SIGNS!!!
\end{subeqnarray}
\begin{subeqnarray}
\theta_3(z,\tau) &\equiv& \sum_{n=-\infty}^{\infty} y^{n} q^{{1\over 2}n^2} \\
                   &=&    1 + 2 \sum_{n=1}^{\infty}
                          q^{{1\over 2}n^2} \cos(2\pi n z)
                          \hspace{3cm}  %% TO LINE UP EQUAL SIGNS!!!
\end{subeqnarray}
\begin{subeqnarray}
\theta_4(z,\tau) &\equiv& \sum_{n=-\infty}^{\infty} (-1)^n y^{n}
                           q^{{1\over 2}n^2} \\
                 &=&       1 + 2 \sum_{n=1}^{\infty}
                           (-1)^n q^{{1\over 2}n^2} \cos(2\pi n z)
                          \hspace{2cm}  %% TO LINE UP EQUAL SIGNS!!!
\end{subeqnarray}
where
\begin{subeqnarray}
q  &=& e^{2 \pi i \tau} \qquad \hbox{with $|q| < 1$} \\
y  &=& e^{2 \pi i z}
\end{subeqnarray}
We sometimes omit the argument $\tau$ when its value is clear from
the context;  in particular, in the present paper we have usually $\tau = i$.
A prime on $\theta_\nu$ indicates the derivative with respect to $z$.

The $\theta$-functions satisfy certain symmetry properties
\begin{subeqnarray}
\theta_1(z\pm1) &=& - \theta_1(z) \\
\theta_2(z\pm1) &=& - \theta_2(z) \\
\theta_3(z\pm1) &=& \phantom{-} \theta_3(z) \\
\theta_4(z\pm1) &=& \phantom{-} \theta_4(z)
\end{subeqnarray}
\begin{subeqnarray}
\theta_1\left(z\pm{1\over2}\right) &=& \pm \theta_2(z) \\
\theta_2\left(z\pm{1\over2}\right) &=& \mp \theta_1(z) \\
\theta_3\left(z\pm{1\over2}\right) &=& \phantom{\pm} \theta_4(z) \\
\theta_4\left(z\pm{1\over2}\right) &=& \phantom{\pm} \theta_3(z)
\end{subeqnarray}
\begin{subeqnarray}
\theta_1(z\pm \tau,\tau) &=& - y^{\mp 1} q^{-1/2} \theta_1(z,\tau) \\
\theta_2(z\pm \tau,\tau) &=& \phantom{-} y^{\mp 1} q^{-1/2} \theta_2(z,\tau) \\
\theta_3(z\pm \tau,\tau) &=& \phantom{-} y^{\mp 1} q^{-1/2} \theta_3(z,\tau) \\
\theta_4(z\pm \tau,\tau) &=& - y^{\mp 1} q^{-1/2} \theta_4(z,\tau)
\end{subeqnarray}
\begin{subeqnarray}
\theta_1\left(z\pm{\tau\over2},\tau\right) &=&
            \pm i y^{\mp 1/2} q^{-1/8} \theta_4(z,\tau) \\
\theta_2\left(z\pm{\tau\over2},\tau\right) &=&
 \phantom{\pm i} y^{\mp 1/2} q^{-1/8} \theta_3(z,\tau) \\
\theta_3\left(z\pm{\tau\over2},\tau\right) &=&
 \phantom{\pm i} y^{\mp 1/2} q^{-1/8} \theta_2(z,\tau) \\
\theta_4\left(z\pm{\tau\over2},\tau\right) &=&
            \pm i y^{\mp 1/2} q^{-1/8} \theta_1(z,\tau)
\end{subeqnarray}

Finally, it is worth noticing that the modulus of a $\theta$-function
satisfies the relation
\be
|\theta_\nu(\pm x_1 \pm i x_2)| = |\theta_\nu(x_1 + i x_2)|
\ee
for $x_1, x_2$ real and $0 \le q < 1$.

The Dedekind $\eta$-function is defined as
\be
\eta(\tau) = q^{1/24} \prod_{n=1}^\infty (1-q^n)
\ee
and it satisfies the relations
\begin{eqnarray}
\theta_2(0,\tau) \theta_3(0,\tau) \theta_4(0,\tau) &=& 2 \eta(\tau)^3 \\[2mm]
\theta^\prime_1(0,\tau) &=& 2 \pi \eta(\tau)^3
   \label{eq_A.13}
\end{eqnarray}
%

%
%  END TEXT SECTIONS
%

\section*{Acknowledgments}

We wish to thank Juan Jes\'us Ruiz-Lorenzo 
for communicating to us his unpublished data;
Chuck Newman for communicating to us his unpublished proof
of a central limit theorem for periodic boxes;
Sergio Caracciolo, Michael Fisher and Andrea Pelissetto for discussions;
and Jae-Kwon Kim, Vladimir Privman and Hubert Saleur for correspondence.
Finally, we wish to thank Carlos Na\'on, Anastasios Petkou,
Andrei Smilga and Aleksandr Sokolov,
who wrote to us after the first version of this paper
was distributed in preprint form,
for pointing out relevant bibliography that we had overlooked.

The authors' research was supported in part
by U.S.\ National Science Foundation grants
PHY-9520978 and PHY-9900769 (A.D.S.)\
and CICyT (Spain) grants PB95-0797 and AEN97-1680 (J.S.).

%%%\newpage
%
\renewcommand{\baselinestretch}{1}
\large\normalsize
%
%
%
%%%%%%%%%%%%   references  %%%%%%%%%%%%%%%%%%%%%%%%
%

\clearpage

%%%%%%%%%%%% START OF TABLES %%%%%%%%%%%%%
\clearpage
\def\kk{\phantom{1}}

%
% TABLE 1
%
%
% Table created by make_table_static_tex from ../table_ising
\begin{table}
\centering 
\begin{tabular}{rcrrr}
\hline\hline\\[-3mm]
$L$ & \multicolumn{1}{c}{MCS} &
        \multicolumn{1}{c}{$\chi$} &
        \multicolumn{1}{c}{$F$} &
        \multicolumn{1}{c}{$\xi$} \\[1mm]
\hline\hline\\[-2mm]
  4 & 2.10  & $    12.1825 \pm \kk   0.0065 $ & $    0.40470 \pm 0.00080 $ & $   3.8146 \pm 0.0051 $ \\  
  6 & 2.70  & $    24.9443 \pm \kk   0.0130 $ & $    0.77280 \pm 0.00130 $ & $   5.5927 \pm 0.0063 $ \\  
  8 & 2.70  & $    41.4214 \pm \kk   0.0228 $ & $    1.25230 \pm 0.00210 $ & $   7.3998 \pm 0.0084 $ \\  
 12 & 3.25  & $    84.3329 \pm \kk   0.0454 $ & $    2.53300 \pm 0.00380 $ & $  10.9783 \pm 0.0114 $ \\  
 16 & 3.25  & $   139.5946 \pm \kk   0.0786 $ & $    4.18240 \pm 0.00630 $ & $  14.5832 \pm 0.0154 $ \\  
 24 & 4.00  & $   284.0239 \pm \kk   0.1525 $ & $    8.49420 \pm 0.01180 $ & $  21.8170 \pm 0.0212 $ \\  
 32 & 4.00  & $   469.7765 \pm \kk   0.2612 $ & $   14.08820 \pm 0.01970 $ & $  29.0118 \pm 0.0288 $ \\  
 48 & 5.00  & $   955.5980 \pm \kk   0.4966 $ & $   28.66410 \pm 0.03650 $ & $  43.4737 \pm 0.0395 $ \\  
 64 & 5.00  & $  1580.9962 \pm \kk   0.8442 $ & $   47.39310 \pm 0.06100 $ & $  57.9660 \pm 0.0535 $ \\  
 96 & 6.40  & $  3214.3979 \pm \kk   1.5807 $ & $   96.39390 \pm 0.11180 $ & $  86.9125 \pm 0.0728 $ \\  
128 & 6.40  & $  5322.9013 \pm \kk   2.6899 $ & $  159.27970 \pm 0.18690 $ & $ 116.0034 \pm 0.0990 $ \\  
192 & 7.10  & $ 10817.0940 \pm \kk   5.3669 $ & $  324.09120 \pm 0.36560 $ & $ 173.8830 \pm 0.1434 $ \\  
256 & 7.10  & $ 17898.9900 \pm \kk   9.0732 $ & $  536.06730 \pm 0.61150 $ & $ 231.8851 \pm 0.1940 $ \\  
512 & 8.10  & $ 60184.2200 \pm   29.9896 $ & $ 1804.17680 \pm 1.96390 $ & $ 463.5381 \pm 0.3745 $ \\  
\\[-3mm]
\hline\hline
\end{tabular}
\caption{Values of the principal static observables for the 2D Ising model
at criticality. For each lattice size $L$ we show the number of measurements
(= Swendsen-Wang iterations after the discard interval) in units
of $10^6$ (MCS), the susceptibility $\chi$, the Fourier-transformed
correlation function $F = \widetilde{G}(2\pi/L,0)$, and the second-moment
correlation length $\xi$.} 
\label{table_statics}
\end{table}
\clearpage

%
% TABLE 2
%
% Table created by make_table_ratios_tex from ../table_ising.mag
%
\begin{table}
\hspace*{-1cm}
\small
\begin{tabular}{rccccc}
\hline\hline\\[-3mm]
$L$ &
\multicolumn{1}{c}{$V_4$}&
\multicolumn{1}{c}{$V_6$}&
\multicolumn{1}{c}{$V_8$}&
\multicolumn{1}{c}{$V_{10}$}&
\multicolumn{1}{c}{$V_{12}$} \\[1mm]
\hline\hline\\[-2mm]
  4 & $ 1.14827 \pm 0.00041 $ & $ 1.38285 \pm 0.00108 $
    &   $  1.7100 \pm  0.0021 $ & $  2.1505 \pm  0.0037 $
    &   $   2.7358 \pm  0.0061 $  \\
  6 & $ 1.15753 \pm 0.00038 $ & $ 1.41690 \pm 0.00102 $
    &   $  1.7941 \pm  0.0021 $ & $  2.3258 \pm  0.0037 $
    &   $   3.0687 \pm  0.0064 $  \\
  8 & $ 1.16042 \pm 0.00039 $ & $ 1.42853 \pm 0.00106 $
    &   $  1.8247 \pm  0.0022 $ & $  2.3930 \pm  0.0040 $
    &   $   3.2026 \pm  0.0069 $  \\
 12 & $ 1.16460 \pm 0.00037 $ & $ 1.44302 \pm 0.00103 $
    &   $  1.8600 \pm  0.0021 $ & $  2.4675 \pm  0.0039 $
    &   $   3.3481 \pm  0.0069 $  \\
 16 & $ 1.16586 \pm 0.00038 $ & $ 1.44774 \pm 0.00106 $
    &   $  1.8721 \pm  0.0022 $ & $  2.4942 \pm  0.0041 $
    &   $   3.4020 \pm  0.0073 $  \\
 24 & $ 1.16672 \pm 0.00036 $ & $ 1.45126 \pm 0.00100 $
    &   $  1.8815 \pm  0.0021 $ & $  2.5151 \pm  0.0039 $
    &   $   3.4447 \pm  0.0069 $  \\
 32 & $ 1.16756 \pm 0.00037 $ & $ 1.45400 \pm 0.00103 $
    &   $  1.8880 \pm  0.0022 $ & $  2.5288 \pm  0.0040 $
    &   $   3.4717 \pm  0.0072 $  \\
 48 & $ 1.16769 \pm 0.00034 $ & $ 1.45475 \pm 0.00094 $
    &   $  1.8903 \pm  0.0020 $ & $  2.5342 \pm  0.0037 $
    &   $   3.4830 \pm  0.0066 $  \\
 64 & $ 1.16777 \pm 0.00034 $ & $ 1.45494 \pm 0.00097 $
    &   $  1.8907 \pm  0.0020 $ & $  2.5353 \pm  0.0038 $
    &   $   3.4854 \pm  0.0068 $  \\
 96 & $ 1.16769 \pm 0.00031 $ & $ 1.45493 \pm 0.00088 $
    &   $  1.8910 \pm  0.0019 $ & $  2.5363 \pm  0.0035 $
    &   $   3.4880 \pm  0.0062 $  \\
128 & $ 1.16763 \pm 0.00032 $ & $ 1.45469 \pm 0.00090 $
    &   $  1.8904 \pm  0.0019 $ & $  2.5351 \pm  0.0036 $
    &   $   3.4857 \pm  0.0063 $  \\
192 & $ 1.16764 \pm 0.00031 $ & $ 1.45474 \pm 0.00087 $
    &   $  1.8906 \pm  0.0018 $ & $  2.5356 \pm  0.0035 $
    &   $   3.4871 \pm  0.0062 $  \\
256 & $ 1.16777 \pm 0.00031 $ & $ 1.45514 \pm 0.00089 $
    &   $  1.8914 \pm  0.0019 $ & $  2.5371 \pm  0.0035 $
    &   $   3.4895 \pm  0.0063 $  \\
512 & $ 1.16782 \pm 0.00030 $ & $ 1.45526 \pm 0.00086 $
    &   $  1.8917 \pm  0.0018 $ & $  2.5376 \pm  0.0034 $
    &   $   3.4906 \pm  0.0061 $  \\
$\infty$&\multicolumn{1}{l}{1.1679229(47)}&
         \multicolumn{1}{l}{1.4556491(72)}&
         \multicolumn{1}{l}{1.89252(18)}&
         \multicolumn{1}{l}{2.53956(34)}& \\
\\[-3mm]
\hline\hline \\[-3mm]
$L$ &
\multicolumn{1}{c}{$V_{14}$}&
\multicolumn{1}{c}{$V_{16}$}&
\multicolumn{1}{c}{$V_{18}$}&
\multicolumn{1}{c}{$V_{20}$}  & \\[1mm]
\hline\hline\\[-2mm]
  4 & $ 3.509 \pm 0.010 $ & $ 4.527 \pm 0.015 $
    &   $ \kk  5.866 \pm  0.022 $ &   $ \kk  7.626 \pm  0.033 $ &   \\
  6 & $ 4.105 \pm 0.010 $ & $ 5.551 \pm 0.017 $
    &   $ \kk  7.575 \pm  0.027 $ &   $  10.414 \pm  0.042 $ &   \\
  8 & $ 4.356 \pm 0.012 $ & $ 6.006 \pm 0.019 $
    &   $ \kk  8.374 \pm  0.031 $ &   $  11.791 \pm  0.049 $ &   \\
 12 & $ 4.628 \pm 0.012 $ & $ 6.496 \pm 0.020 $
    &   $ \kk  9.241 \pm  0.033 $ &   $  13.300 \pm  0.054 $ &   \\
 16 & $ 4.731 \pm 0.012 $ & $ 6.688 \pm 0.021 $
    &   $ \kk  9.589 \pm  0.035 $ &   $  13.921 \pm  0.058 $ &   \\
 24 & $ 4.814 \pm 0.012 $ & $ 6.843 \pm 0.020 $
    &   $ \kk  9.874 \pm  0.034 $ &   $  14.436 \pm  0.057 $ &   \\
 32 & $ 4.864 \pm 0.012 $ & $ 6.936 \pm 0.021 $
    &   $  10.040 \pm  0.036 $ &   $  14.732 \pm  0.060 $ &   \\
 48 & $ 4.887 \pm 0.012 $ & $ 6.979 \pm 0.020 $
    &   $  10.123 \pm  0.033 $ &   $  14.883 \pm  0.056 $ &   \\
 64 & $ 4.892 \pm 0.012 $ & $ 6.990 \pm 0.020 $
    &   $  10.143 \pm  0.034 $ &   $  14.922 \pm  0.058 $ &   \\
 96 & $ 4.898 \pm 0.011 $ & $ 7.001 \pm 0.019 $
    &   $  10.165 \pm  0.031 $ &   $  14.965 \pm  0.053 $ &   \\
128 & $ 4.894 \pm 0.011 $ & $ 6.995 \pm 0.019 $
    &   $  10.154 \pm  0.032 $ &   $  14.947 \pm  0.054 $ &   \\
192 & $ 4.897 \pm 0.011 $ & $ 7.001 \pm 0.018 $
    &   $  10.166 \pm  0.031 $ &   $  14.970 \pm  0.053 $ &   \\
256 & $ 4.901 \pm 0.011 $ & $ 7.006 \pm 0.019 $
    &   $  10.175 \pm  0.032 $ &   $  14.982 \pm  0.054 $ &   \\
512 & $ 4.903 \pm 0.011 $ & $ 7.011 \pm 0.018 $
    &   $  10.183 \pm  0.031 $ &   $  14.997 \pm  0.052 $ &   \\
\\[-3mm]
\hline\hline
\end{tabular}
\caption{Values of the ratios
$V_{2n} =\<{\cal M}^{2n}\>/\<{\cal M}^2\>^n$ for the 2D Ising
model at criticality, as a function of the lattice size $L$. The row
$L=\infty$ shows the theoretical predictions
\protect\reff{V4_exact}/\protect\reff{V6_exact}/%
\protect\reff{V8_exact}/\protect\reff{V10_exact}
for $V_4$, $V_6$, $V_8$ and $V_{10}$, respectively;
they are exact except for a numerical integration,
the error bars of which are given in parentheses.}
\label{table_ratios}
\end{table}
%\clearpage

%
% TABLE 3 (XI DATA FROM COMPLUTENSE GROUP)
%
\begin{table}
\centering
\begin{tabular}{rcl}
\hline\hline\\[-3mm]
$L$ & \multicolumn{1}{c}{EIM} & \multicolumn{1}{c}{$\xi$} \\[1mm]
\hline\hline\\[-2mm]
12  & 0.4 & $\kk 10.976    \pm 0.015    $\\
16  & 0.4 & $\kk 14.575    \pm 0.019    $\\
24  & 0.6 & $\kk 21.791    \pm 0.026    $\\
32  & 0.4 & $\kk 29.089    \pm 0.038    $\\
48  & 0.6 & $\kk 43.448    \pm 0.043    $\\
64  & 0.6 & $\kk 57.877    \pm 0.056    $\\
96  & 0.6 & $\kk 86.87\kk  \pm 0.11\kk  $\\
128 & 0.6 & $115.87\kk     \pm 0.12\kk  $\\
192 & 0.5 & $174.06\kk     \pm 0.21\kk  $\\
256 & 0.6 & $231.84\kk     \pm 0.31\kk  $\\
512 & 0.7 & $464.8\kk\kk   \pm 0.5\kk\kk$\\
\\[-3mm]
\hline\hline
\end{tabular}
\caption{Values of the correlation length $\xi$ for the 2D Ising model
at $\beta=\beta_c$ obtained by Ballesteros {\em et al}.\
\protect\cite{complutense}. For each lattice size $L$ we also show the
number of ``effectively independent measurements''
in units of $10^6$ (EIM).}
\label{table_xi_JJ}
\end{table}
%\clearpage

%
% TABLE 4 (MERGED XI DATA)
%
\begin{table}
\centering
\begin{tabular}{rrrr}
\hline\hline\\[-3mm]
$L$ & \multicolumn{1}{c}{$\xi$} & \multicolumn{1}{c}{$\xi/L$} &
                                  \multicolumn{1}{c}{$\xi'/L$} \\[1mm]
\hline\hline\\[-2mm]
   4& $    3.8146 \pm  0.0051 $&$0.95365 \pm 0.00128$&$0.85859 \pm 0.00115$\\
   6& $    5.5927 \pm  0.0063 $&$0.93212 \pm 0.00105$&$0.89011 \pm 0.00100$\\
   8& $    7.3998 \pm  0.0084 $&$0.92497 \pm 0.00105$&$0.90138 \pm 0.00102$\\
  12& $   10.9775 \pm  0.0091 $&$0.91479 \pm 0.00076$&$0.90437 \pm 0.00075$\\
  16& $   14.5799 \pm  0.0120 $&$0.91125 \pm 0.00075$&$0.90540 \pm 0.00074$\\
  24& $   21.8066 \pm  0.0164 $&$0.90861 \pm 0.00068$&$0.90602 \pm 0.00068$\\
  32& $   29.0400 \pm  0.0230 $&$0.90750 \pm 0.00072$&$0.90604 \pm 0.00072$\\
  48& $   43.4619 \pm  0.0291 $&$0.90546 \pm 0.00061$&$0.90481 \pm 0.00061$\\
  64& $   57.9235 \pm  0.0387 $&$0.90506 \pm 0.00060$&$0.90469 \pm 0.00060$\\
  96& $   86.8996 \pm  0.0607 $&$0.90520 \pm 0.00063$&$0.90504 \pm 0.00063$\\
 128& $  115.9494 \pm  0.0764 $&$0.90585 \pm 0.00060$&$0.90576 \pm 0.00060$\\
 192& $  173.9393 \pm  0.1184 $&$0.90593 \pm 0.00062$&$0.90589 \pm 0.00062$\\
 256& $  231.8724 \pm  0.1645 $&$0.90575 \pm 0.00064$&$0.90573 \pm 0.00064$\\
 512& $  463.9916 \pm  0.2997 $&$0.90623 \pm 0.00059$&$0.90623 \pm 0.00059$\\
$\infty$&                      &\multicolumn{1}{l}{$0.9050488$}
                               &\multicolumn{1}{l}{$0.9050488$} \\
\\[-3mm]
\hline\hline
\end{tabular}
\caption{Values of the correlation length $\xi$ for the 2D Ising model at
$\beta=\beta_c$ coming from merging our data
(see Table~\protect\ref{table_statics}) with that of
Ballesteros {\em et al}.\ \protect\cite{complutense} (see
Table~\protect\ref{table_xi_JJ}).
The second column shows the ratio $\xi/L$, and the last column shows the
ratio $\xi'/L$.
The last row ($L=\infty$) shows the theoretical prediction
\protect\reff{xstar_theor} for the infinite-volume limit of the ratios
$\xi/L$ and $\xi'/L$.
}
\label{table_xi_merged}
\end{table}
\clearpage

%
% TABLE 5
%
% Table created by make_table_ratios_fit_tex from ../fit_ising_Tc_ratios.all
%
\begin{table}
\hspace*{-1.5cm}
\begin{tabular}{rcrrlrccr}
\hline\hline\\[-3mm]
$2n$ & Type & \multicolumn{1}{c}{$V_{2n}^\infty$} &
\multicolumn{1}{c}{$B$}&\multicolumn{1}{c}{$\Delta$} &
\multicolumn{1}{c}{$L_{min}$} & $\chi^2$ & $DF$ &
\multicolumn{1}{c}{level}\\[1mm]
\hline\hline\\[-2mm]
 4 & C & $  1.16770\kk\kk \pm 0.00011\kk\kk $ & & &   32 & 0.48 & 7 & 100\% \\
   & P & $  1.16777\kk\kk \pm 0.00013\kk\kk $
   & $    -0.477 \pm \kk    0.228$ & $    2.007 \pm  0.223$
   &   8 & 1.53 & 9 & 100\% \\
%  &P${}'$ & $1.1679227 \pm 0.0000047$ & $-0.361 \pm \kk 0.140$ &
%      $1.866 \pm 0.174$ & 8 & 2.91 & 10 & 98\% \\
  & T & $1.1679229 \pm 0.0000047$ & & & & & &  \\
\hline\\[-2mm]
 6 & C & $  1.45484\kk\kk \pm 0.00032\kk\kk $ & & &   32 & 1.09 & 7 &  99\% \\
   & P & $  1.45517\kk\kk \pm 0.00037\kk\kk $
   & $    -1.387 \pm \kk    0.476$ & $    1.901 \pm  0.160$
   &   8 & 1.66 & 9 & 100\% \\
%  &P${}'$ & $1.4556489 \pm 0.0000072$ & $-1.111 \pm \kk 0.309$ &
%      $1.788 \pm 0.124$ & 8 & 3.24 & 10 & 98\% \\
  & T & $1.4556491 \pm 0.0000072$ & & & & & & \\
\hline\\[-2mm]
 8 & C & $  1.89090\kk\kk \pm 0.00071\kk\kk $ & & &   48 & 0.48 & 6 & 100\% \\
   & P & $  1.89163\kk\kk \pm 0.00079\kk\kk $
   & $    -3.037 \pm \kk    0.827$ & $    1.834 \pm  0.127$
   &   8 & 1.86 & 9 &  99\% \\
%  &P${}'$ & $1.89248\kk\kk \pm 0.00018\kk\kk$ & $-2.612 \pm \kk 0.589$ &
%      $1.757 \pm 0.101$ & 8 & 3.03 & 10 & 98\% \\
  & T & $1.89252\kk\kk \pm 0.00018\kk\kk$ & & & & & & \\
\hline\\[-2mm]
10 & C & $  2.53593\kk\kk \pm 0.00135\kk\kk $ & & &   48 & 0.69 & 6 &  99\% \\
   & P & $  2.53769\kk\kk \pm 0.00151\kk\kk $
   & $    -5.915 \pm \kk    1.341$ & $    1.784 \pm  0.106$
   &   8 & 2.06 & 9 &  99\% \\
%  &P${}'$ & $2.53947\kk\kk \pm 0.00033\kk\kk$ & $-5.136 \pm \kk 0.960$ &
%      $1.712 \pm 0.083$ & 8 & 3.49 & 10 & 97\% \\
  & T & $2.53956\kk\kk \pm 0.00034\kk\kk$ & & & & & &  \\
\hline\\[-2mm]
12 & C & $  3.48720\kk\kk \pm 0.00241\kk\kk $ & & &   48 & 0.99 & 6 &  99\% \\
   & P & $  3.49106\kk\kk \pm 0.00273\kk\kk $
   & $   -10.819 \pm \kk    2.112$ & $    1.742 \pm  0.091$
   &   8 & 2.29 & 9 &  99\% \\
\hline\\[-2mm]
14 & C & $  4.89621\kk\kk \pm 0.00418\kk\kk $ & & &   48 & 1.37 & 6 &  97\% \\
   & P & $  4.90419\kk\kk \pm 0.00479\kk\kk $
   & $   -19.019 \pm \kk    3.264$ & $    1.705 \pm  0.080$
   &   8 & 2.56 & 9 &  98\% \\
\hline\\[-2mm]
16 & C & $  6.99812\kk\kk \pm 0.00716\kk\kk $ & & &   48 & 1.83 & 6 &  93\% \\
   & P & $  7.01407\kk\kk \pm 0.00827\kk\kk $
   & $   -32.544 \pm \kk    4.984$ & $    1.670 \pm  0.072$
   &   8 & 2.86 & 9 &  97\% \\
\hline\\[-2mm]
18 & C & $ 10.16503\kk\kk \pm 0.01303\kk\kk $ & & &   64 & 0.97 & 5 &  97\% \\
   & P & $ 10.19047\kk\kk \pm 0.01416\kk\kk $
   & $   -54.624 \pm \kk    7.553$ & $    1.635 \pm  0.065$
   &   8 & 3.23 & 9 &  95\% \\
\hline\\[-2mm]
20 & C & $ 14.96506\kk\kk \pm 0.02199\kk\kk $ & & &   64 & 1.16 & 5 &  95\% \\
   & P & $ 15.01380\kk\kk \pm 0.02411\kk\kk $
   & $   -90.377 \pm    11.374$ & $    1.601 \pm  0.059$
   &   8 & 3.69 & 9 &  93\% \\
\hline\hline
\end{tabular}
\caption{
   Values of the infinite-volume-limit ratios
   $V_{2n} = \<{\cal M}^{2n}\>/\<{\cal M}^2\>^n$
   for the 2D Ising model at criticality.
   For each $n$ we show the results of two different types of fits:
   to a constant $V_{2n} = V_{2n}^\infty$ (C),
   and to a constant plus a power-law correction-to-scaling term
   $V_{2n} = V_{2n}^\infty + B_{2n} L^{-\Delta}$ (P).
   We also show, for comparison, the theoretical prediction itself (T) for
   $2n =4,6,8,10$.
   The values of $L_{min}$, $\chi^2$, the number of degrees of freedom (DF)
   and the confidence level are also shown.
}
\label{table_fit_ratios}
\end{table}
\clearpage

%%%%%%%%%%% START OF FIGURES %%%%%%%%%%%%%
%
% FIGURE 1
%

\end{document}